\documentstyle[12pt,frascatiphys,epsf]{article}
\newcommand{\gev}{\, {\rm GeV}}
\newcommand{\mev}{\, {\rm MeV}}
\newcommand{\im}{{\rm Im}}

\newcommand{\Gab}{\bar \Gamma}
\newcommand{\co}{\; \; ,}
\newcommand{\fs}{\; \; .}
\newcommand{\nn}{\nonumber \\}
\newcommand{\dg}{\dagger}
\newcommand{\be}{\begin{equation}}
\newcommand{\ee}{\end{equation}}
\newcommand{\bea}{\begin{eqnarray}}
\newcommand{\eea}{\end{eqnarray}}
\newcommand{\beq}{\begin{equation}}
\newcommand{\eeq}{\end{equation}}
\newcommand{\ba}{\begin{array}}
\newcommand{\ea}{\end{array}}
\newcommand{\beqa}{\begin{eqnarray}}
\newcommand{\eeqa}{\end{eqnarray}}
\newcommand{\dis}{\displaystyle}
\newcommand{\cL}{{\cal L}}

\newcommand{\cA}{{\cal A}}
\newcommand{\cT}{{\cal T}}
\newcommand{\cO}{{\cal O}}
\newcommand{\cB}{{\cal B}}
\newcommand{\da}{^\dagger}
\newcommand{\no}{\nonumber}

\newcommand{\gsim}{\stackrel{>}{_\sim}}
\newcommand{\Gto}{\stackrel{G}{\longrightarrow}}     
\newcommand{\Hto}{\stackrel{H}{\longrightarrow}}  
\newcommand{\ket}[1]{\vert {#1} \rangle}
\newcommand{\bra}[1]{\langle {#1}}
\newcommand{\Ko}{{K}^0}
\newcommand{\Kob}{\bar{K}^0} 
\newcommand{\la}{\langle}
\newcommand{\ra}{\rangle}
\newcommand{\DD}{\bigtriangledown}

\def\ap#1#2#3{     {\it Ann. Phys.  }{\bf #1} (#2) #3}
\def\arnps#1#2#3{  {\it Annu. Rev. Nucl. Part. Sci. }{\bf #1} (#2) #3}
\def\npb#1#2#3{    {\it Nucl. Phys. }{\bf B #1} (#2) #3}
\def\plb#1#2#3{    {\it Phys. Lett. }{\bf B #1} (#2) #3}
\def\pr#1#2#3{     {\it Phys. Rev. }{\bf   #1} (#2) #3}
\def\prd#1#2#3{    {\it Phys. Rev. }{\bf D #1} (#2) #3}
\def\prb#1#2#3{    {\it Phys. Rev. }{\bf B #1} (#2) #3}

\def\prl#1#2#3{    {\it Phys. Rev. Lett. }{\bf #1} (#2) #3}
\def\ptp#1#2#3{    {\it Prog. Theor. Phys. }{\bf #1} (#2) #3}

\def\ppnp#1#2#3{   {\it Prog. Part. Nucl. Phys. }{\bf #1} (#2) #3}
\def\rmp#1#2#3{    {\it Rev. Mod. Phys. }{\bf #1} (#2) #3}
\def\zpc#1#2#3{    {\it Z. Phys. }{\bf C #1} (#2) #3}
\def\ijmpa#1#2#3{  {\it Int. J. Mod. Phys. }{\bf A #1} (#2) #3}

\def\nc#1#2#3{     {\it Nuovo Cim. }{\bf #1} (#2) #3}

\def\jhep#1#2#3{   {\it JHEP  }{\bf #1} (#2) #3} 
\def\physica#1#2#3{{\it Physica }{\bf A #1} (#2) #3}
\def\ncim#1#2#3{   {\it Nuovo Cim. }{\bf #1} (#2) #3}

\begin{document}              


\begin{flushright}
ZU-TH  04/01     \\
CERN-TH/2001-019 \\
hep-ph/0101264 
\end{flushright}

\vskip 2 cm 

\begin{center}

{\bf AN INTRODUCTION TO CHPT} \\ [10pt]

Gilberto Colangelo$^1$ and Gino Isidori$^2$ \\ [5pt]
{\em ${}^{1)}$ Institut f{\"u}r Theoretische 
Physik der Universit{\"a}t Z\"urich,} \\
{\em Winterthurerstr. 190, CH--8057 Zurich--Irchel, Switzerland}\\ [3pt]
{\em ${}^{2)}$ Theory Division, CERN, CH-1211 Geneva 23, 
Switzerland~\footnote{On leave from INFN, Laboratori Nazionali di 
Frascati, Via Enrico Fermi 40, I-00044 Frascati (Rome), Italy.}}

\vskip 2 cm 
{\bf Abstract} \\
\end{center}

\noindent
These lectures provide an elementary introduction to 
Chiral Perturbation Theory, focused on the 
sector of pseudoscalar meson interactions.
Basic concepts and technical methods of this 
approach are discussed on general grounds
and with the help of a few specific examples.

\begin{center} 
\vskip 4 cm
Lectures given at the \\ [3pt]
{\bf 2000 LNF Spring School} \\
{\em in Nuclear, Subnuclear and Astroparticle Physics} \\ [3pt]
Frascati, Italy, 15--20 May 2000

\end{center}
\thispagestyle{empty}
\setcounter{page}{0}
\newpage

\centerline{\bf AN INTRODUCTION TO CHPT}
\vskip 1 cm 
\tableofcontents

\newpage


\title{AN INTRODUCTION TO CHPT}
\author{Gilberto Colangelo$^1$ and Gino Isidori$^2$  \\ [5pt]
{\em ${}^{1)}$ Institut f{\"u}r Theoretische 
Physik der Universit{\"a}t Z{\"u}rich,} \\
{\em Winterthurerstr. 190, CH--8057 Z{\"u}rich--Irchel, Switzerland}\\ [3pt]
{\em ${}^{2)}$ Theory Division, CERN, CH-1211 Geneva 23, 
Switzerland}\\ 
}



\baselineskip=17pt
\section*{Prologue}
\addcontentsline{toc}{section}{Prologue}

Chiral Perturbation Theory (CHPT) is nothing but the low-energy limit of
the Standard Model (SM) or, to be more precise, the effective quantum field
theory describing hadronic interactions according to the SM, below the
breaking scale of chiral symmetry ($E\ll \Lambda_\chi \sim 1$~GeV). This
theory, founded a long time ago by the pioneering works of
Weinberg,\cite{Weinberg79} and Gasser and Leutwyler,\cite{GL1} is nowadays a
rather mature subject. On one side several two-loop calculations have
been performed in the purely mesonic sector, reaching, in some cases, a
very high degree of precision. On the other side the original
formulation has been successfully extended in several directions,
including, for instance, heavy quark fields, bound-state dynamics,
non-zero temperature effects, etc.

A complete overview of the subject would be a tremendous task, and we could
certainly not provide it in the four hours that we were given for these
lectures. The interested reader is referred to some excellent 
reviews\cite{reviews,handbook,Ecker} for a broader survey
of the subject. The purpose of these lectures is to provide a basic
introduction to CHPT. We will therefore restrict our attention to the
simplest case, namely the mesonic sector, discussing in detail the
determination of the effective Lagrangians and the calculation of a few
specific quantities.

The lectures are organized as follows: motivations, basic principles and
the tree-level structure of CHPT are presented in the first
lecture. Various aspects of loop calculations are discussed in the following
two lectures. In particular, the renormalization at the one-loop level is
presented in the second lecture, whereas the third lecture is devoted to
studying the properties of unitarity and analyticity, and to showing how
dispersion relations can be combined with the chiral expansion. Finally,
the issue of non-leptonic weak interactions is introduced in the last
lecture. 

\section{Generalities and lowest-order Lagrangians}

\subsection{Effective quantum field theories}

Within the Standard Model the interactions between quarks and gluons, ruled
by Quantum Chromo Dynamics (QCD), are highly non-perturbative at energies
below the breaking scale of chiral symmetry. This makes very difficult any
description of the low-energy hadronic world in terms of partonic degrees
of freedom. On the other hand, the spectrum of the theory is rather simple
at low energies, containing only the octet of light pseudoscalar mesons:
$\pi$, $K$ and $\eta$.  Experimentally we also know that, at very low
energies, these pseudoscalar mesons interact weakly, both among themselves
and with nucleons.  It is then reasonable to expect that QCD can be treated
perturbatively even at low energies, provided a suitable transformation of
degrees of freedom is performed. This is exactly the goal of Chiral
Perturbation Theory, where the pseudoscalar mesons are assumed to be
fundamental degrees of freedom.

Having an intrinsic energy limitation and being the low-energy limit of a
more fundamental theory, CHPT is a typical example of {\em effective}
quantum field theory (EQFT), a widely used tool in modern physics.  The
basic principle of any EQFT is that, in a given energy range, only few
degrees of freedom are relevant and need to be described by dynamical
fields.  The remaining degrees of freedom of the more general theory can be
integrated out, leading to effects that are encoded in the coefficients of
appropriate local operators.

Assuming this general point of view, all known quantum field theories,
including the Standard Model, can be considered as effective. An
important distinction, however, is provided by the degree of
renormalizability. In general the requirement of renormalizability,
understood in the classical sense, is not mandatory within an
EQFT.\cite{Georgi} Indeed if the theory is meant to be valid only for
energies below a given cut-off $\Lambda$, and we perform an expansion of
the physical amplitudes in powers of $E/\Lambda$, we can impose the weaker
condition that, for any $n > 0$, the number of counterterms needed to
regularize the amplitudes and contributing at $\cO[(E/\Lambda)^n]$ is
finite. This condition is sufficient to ensure a predictive power to the
theory for $E < \Lambda$.  In other words, within an EQFT we only need
renormalizability order by order in the energy expansion.
The classical requirement of renormalizability, is a stronger constraint,
which prevents a naive derivation of the intrinsic cut-off of the
theory. The SM belongs to this restricted subclass of EQFTs without an
intrinsic cut-off, whereas CHPT belongs to the more general case.

\subsection{Chiral symmetry}
\label{sect:chsymm}

Neglecting light-quark masses, the QCD Lagrangian can be written as
\beq
\cL^{(0)}_{QCD}=\sum_{q=u,d,s} {\bar q} \gamma^\mu \left(
i\partial_\mu - {g_s} {\lambda_a \over 2} G_\mu^a  \right) q
 -{1\over 4} G^a_{\mu\nu} G^{a\mu\nu}+\cO(\mbox{\rm heavy quarks}).
\eeq
Apart from the $SU(3)_C$ local invariance,
$\cL^{(0)}_{QCD}$ has a global invariance under the group
$SU(N_f)_L\times SU(N_f)_R\times U(1)_V\times U(1)_A$,
where $N_f$ is the number of massless quarks
($N_f=3$ in the case considered above). The 
$U(1)_V$ symmetry, which survives also in the case of non-vanishing
quark masses, is exactly conserved and its generator is the 
baryonic number. The $U(1)_A$  symmetry is 
explicitly broken at the quantum level by the Abelian 
anomaly.\cite{Hooft76} Finally 
$G=SU(3)_L\times SU(3)_R$ is the group of chiral transformations:
\beq
\psi_{_{L,R}} \Gto g_{_{L,R}} \psi_{_{L,R}}, 
\qquad \mbox{\rm where}\qquad
\psi=\left( \ba{c} u \\ d \\ s \ea\right)\qquad \mbox{\rm and}\qquad
g_{_{L,R}}\in G~.
\eeq

If the operator $\bar{\psi}\psi$ has a non-vanishing expectation value in
the ground state ($\bra{0}|\bar{\psi}\psi\ket{0}\not=0$), or in the presence of
a non-vanishing quark condensate, chiral symmetry is spontaneously
broken. The subgroup that remains unbroken after the breaking of $G$ is
$H=SU(3)_V\equiv SU(3)_{L+R}$, the famous $SU(3)$ of the {\em eightfold
way},\cite{Gellmann} or, in the limit where only two quarks are kept
massless, the $SU(2)$ group of isospin transformations.

The fundamental idea of CHPT is that, in the chiral limit ($m_u=m_d=m_s=0$
or, eventually, $m_u=m_d=0$), the light pseudoscalar mesons are the
Goldstone bosons generated by the spontaneous breaking of $G$ into $H$ (in
the $SU(3)$ case, $m_u=m_d=m_s=0$, the full octet of pseudoscalar mesons is
identified with Goldstone fields, whereas in the $SU(2)$ case, $m_u=m_d=0$,
only the three pions are). These light particles have the correct quantum
numbers to be associated with the generators of $G/H$, as required by the
Goldstone theorem.\cite{Goldstone} Moreover, since Goldstone fields can
always be redefined so that they interact only through derivative
couplings,\cite{Goldstone} this hypothesis justifies the soft behaviour of
pseudoscalar interactions at low energies. If pseudoscalar mesons were
effectively Goldstone bosons, they would had been massless. This does not
happen in the real world, owing to the light-quark mass terms, which
explicitly break $G$. Since $m_{u,d,s} < \Lambda_\chi$, it is natural to
expect that these breaking terms can be treated as small perturbations. The
fact that pseudoscalar-meson masses are much smaller than the typical
hadronic scale indicates that also this hypothesis is reasonable.  Clearly
this approximation works much better in the $SU(2)$ case ($M^2_\pi/M^2_\rho
\sim 0.03$) than in the $SU(3)$ one ($M^2_\pi/M^2_\rho \sim 0.4$).
Summarizing, the two basic assumptions of CHPT are the following:
\begin{enumerate}
\item {\em In the chiral limit the $SU(3)_L\times SU(3)_R$
symmetry of the QCD Lagrangian is spontaneously broken into 
$SU(3)_{L+R}$ and the pseudoscalar meson fields can be 
identified with the corresponding Goldstone bosons.}
\item {\em The mass terms of light quarks 
can be treated as small perturbations around the chiral limit.}  
\end{enumerate}
According to these hypotheses, in order to describe the QCD interactions of
pseudoscalar mesons it is necessary to construct, in terms of
Goldstone-boson fields, the most general Lagrangian invariant under $G$ and
add to it the explicit breaking terms (light-quark masses) that transform
linearly under $G$.\cite{Weinberg79} The Lagrangian built in this way
necessarily contains an infinite number of operators. Nevertheless, as
anticipated, only a finite number of operators contribute at
$\cO[(E/\Lambda_\chi)^n]$). Therefore the theory can reach an arbitrary
degree of precision for processes occurring at $E<\Lambda_\chi$, provided a
sufficient (but finite) number of couplings is fixed by experimental data.

\subsection{Non-linear realization of chiral symmetry}
\label{sez:nonlinear}

Denoting by $V_i$ the generators of $H$ 
and by $A_i$ the remaining generators of $G$,
any element of $G$ can be unambiguously decomposed 
as $g=e^{\xi_i  A_i} e^{\eta_i  V_i}$. 
The Goldstone-boson fields are associated to the 
coordinates $\xi_i$ of the coset space $G/H$.
To understand how these transform under $G$
we consider the action of a generic 
element $g \in G$ on $u(\xi_i)=e^{\xi_i  A_i}$:
\beq
g \in G \qquad\qquad  g e^{\xi_i  A_i} = e^{\xi^\prime_i (g, \xi) A_i} e^{\eta^\prime_i (g, \xi) V_i}~.
\eeq
The transformation $u(\xi_i) \Gto u(\xi_i^\prime)$
provides a non-linear realization of the group $G$.\cite{CWZ}  
This realization is not linear since $V_i$'s and 
$A_i$'s do not commute ($[V_i,A_j] \sim A_k$); however, it 
becomes linear if restricted to the subgroup $H$:
\beq
h_0 = e^{\eta^0_i  V_i} \in H~, \qquad\qquad  h_0 e^{\xi_i  A_i} =  \left[
e^{\eta^0_i  V_i}  e^{\xi_i  A_i} e^{- \eta^0_i  V_i} \right] 
e^{\eta_i^0 V_i}~.
\eeq
As shown by Callan, Coleman, Wess and Zumino,\cite{CWZ}
this non-linear realization provides the most general tool 
to construct operators that transform linearly under $G$ 
(or invariant operators) in terms of 
the Goldstone-boson fields generated by the spontaneous 
breakdown of $G$ into $H$.

In the specific case of chiral symmetry an important 
additional information is provided by the automorphism 
induced on $G$ by parity 
[$P: (A_i,V_i) \to (-A_i,V_i)$]. This implies  
that if $g_R : u(\xi_i) \to u(\xi_i^\prime)$, then 
$g_L: u(-\xi_i) = u(\xi_i)\da  \to u(\xi_i^\prime)\da$. 
We can therefore write 
\beqa 
u(\xi_i) &\Gto& g_{_R} u(\xi_i) h^{-1}(g,\xi_i) =
h(g,\xi_i) u(\xi_i) g_{_L}^{-1}~, \no\\
u(\xi_i)\da &\Gto& g_{_L} u(\xi_i)\da h^{-1}(g,\xi_i) =
h(g,\xi_i) u(\xi_i)\da g_{_R}^{-1}, 
\label{eq:u}
\eeqa
where $h(g,\xi_i) = e^{\eta^\prime(g,\xi_i) V}$. 
At this point we have all the ingredients to build 
generic operators transforming linearly under $G$,
starting from their projection into $H$.
For instance, given a generic field $\Psi$ transforming 
linearly under $H$ as $\Psi \Hto e^{\eta_i V_i}
 \Psi e^{-\eta_i V_i}$,
in the non-linear realization of $G$ we find
\beq
\Psi \Gto  h(g,\xi_i) \Psi  h^{-1}(g,\xi_i)~,
\label{eq:h}
\eeq
thus any product of the type 
$(u,u\da) \times  \Psi \times (u,u\da)$
transforms linearly under $G$:
\beqa 
u    \Psi u\da &\Gto&  g_{_R} (u     \Psi u\da) g_{_R}^{-1}, \no\\
u\da \Psi u    &\Gto&  g_{_L} (u\da  \Psi u   ) g_{_L}^{-1}, \no\\
u    \Psi u    &\Gto&  g_{_R} (u  \Psi    u   ) g_{_L}^{-1}, \no\\
u\da \Psi u\da &\Gto&  g_{_L} (u\da \Psi  u\da) g_{_R}^{-1}.  
\label{eq:Oplin}
\eeqa  
The above procedure can be generalized to the case of 
fields $\Psi^\prime$ belonging to different representations of $H$.
Furthermore, starting from the derivative of $u$ and 
$u\da$ we can define the following operators:
\beqa       
u_\mu &=& i(u\da \partial_\mu u  - u \partial_\mu u\da) = iu\da\partial_\mu Uu\da
= u_\mu\da~,    \\
\Gamma_\mu &=& {1\over 2}(u\da \partial_\mu u  + u \partial_\mu u\da) 
= -\Gamma_\mu\da~.
\eeqa
It is easy to verify that both $u_\mu$ and the covariant derivative of $\Psi$, 
\beq
\DD_\mu \Psi = \partial_\mu  \Psi - [\Gamma_\mu, \Psi]~,
\eeq  
transform according to Eq.~(\ref{eq:h}). 
Thus, similarly to Eq.~(\ref{eq:Oplin}), 
with appropriate contractions with $u$ and $u\da$ 
we can construct operators transforming linearly 
under $G$ also starting from $u_\mu$ and $\DD_\mu \Psi$.
Concerning the construction of terms invariant
under $G$, we note that given a generic operator 
$Q$ transforming as $Q\Gto g_L Q g_L^{-1}$ or  
$Q\Gto g_R Q g_R^{-1}$,
an invariant term is obtained by the trace of $Q$ 
in flavour space, denoted in the following by  $\la Q\ra$.

The choice of coordinates in the coset space $G/H$ 
is not unique; however, in any given set of 
coordinates we can introduce a field $u(\xi_i)$
transforming as in Eq.~(\ref{eq:u}).\cite{CWZ} 
The freedom in the choice of coordinates implies 
that the parametrization of $u$ in terms of the 
pseudoscalar meson fields is not unique. 
In the following we shall adopt the exponential 
parametrization in the $3 \times 3$ flavour space, 
defined by             
\beqa
u^2 = U &=& e^{i\sqrt{2}\Phi/F}, \no\\
  \Phi &=& {1\over \sqrt{2} } \sum_i\lambda_i\phi^i 
= \left[ \ba{ccc}\dis{\pi^0\over \sqrt{2}}+\dis{\eta_8 \over \sqrt{6}}&\pi^+ &
 K^+ \\ 
\pi^- & -\dis{\pi^0\over\sqrt{2}}+{\eta_8 \over \sqrt{6}} & \Ko \\ 
K^- & \Kob & -\dis{2\eta_8 \over \sqrt{6}} 
\ea \right]~, \label{eq:Phi}
\eeqa
where $\eta_8$ denotes the octet component of the $\eta$ meson.
The parameter $F$ appearing in Eq.~(\ref{eq:Phi}) 
is a dimensional constant (dim$[F]$~=~dim$[\Phi]$)
that, as we shall see in the following, can be related to 
the decay constant of pseudoscalar mesons.

\subsection{Lowest-order (strong) Lagrangian}

In the absence of external fields it is impossible to build 
non-trivial invariant operators in terms of 
$u$ and $u\da$ only, without their derivatives:
it is necessary to have at least two derivatives acting on
$u$ or $u\da$ in order to build a non-trivial 
structure invariant under both chiral and Lorenz 
symmetries. If only two derivatives 
are considered this structure is unique:
\beq
\la u_\mu u^\mu \ra=\la \partial_\mu U \partial^\mu U\da \ra~.
\eeq
Fixing the coupling constant of the  operator $\la u_\mu u^\mu \ra$
so as to reproduce the correct kinetic term 
of spinless fields, leads to
\beq
{\widetilde \cL}_S^{(2)} ~=~ {F^2 \over 4} \la \partial_\mu U \partial^\mu U\da \ra~.
\eeq
This Lagrangian is the chiral realization of
$\cL^{(0)}_{QCD}$ at the lowest 
order in the derivative expansion. 

Expanding ${\widetilde \cL}_S^{(2)}$ in powers of 
$\Phi$ leads to an infinite series of operators, 
whose couplings are all determined in terms of $F$: 
\beq
{\widetilde \cL}_S^{(2)} = {F^2 \over 4} \la \partial_\mu U \partial^\mu U\da \ra \\ 
= {1\over 2} \la \partial_\mu \Phi  \partial^\mu \Phi   \ra 
+ {1 \over 6F^2} \la \left[  \partial_\mu \Phi, \Phi \right]  
\left[  \partial^\mu \Phi, \Phi \right] \ra + \cO(\Phi^6)~.
\label{eq:L2exp}
\eeq
From ${\widetilde \cL}_S^{(2)}$ one can therefore determine the 
amplitude for any process of the type 
$\pi_1 \pi_2 \ldots \pi_n \to \pi^\prime_1 \pi^\prime_2 \ldots \pi^\prime_m$
in the chiral limit. For example, considering the 
second term in Eq.~(\ref{eq:L2exp}) it is easy to show that 
\beq
\left. \cA (\pi^+ \pi^0 \to  \pi^+ \pi^0 )\right|_{[m_q = 0]}
~=~ \frac{(p_+ - p_+^\prime)^2}{F^2} + \cO(p^4)~.
\eeq

To parametrize the breaking terms induced by quark 
masses, and also to generate in a systematic way  
the Green functions of quark currents, it 
is convenient to insert appropriate external
sources both in the QCD Lagrangian and in its 
chiral realization.
Following Gasser and Leutwyler,\cite{GL1}  
we introduce the sources $v_\mu$, $a_\mu$, $s$ and $p$, 
which transform as
\beqa
r_\mu = v_\mu + a_\mu  &\Gto &  g_{_R} r_\mu g_{_R}^{-1}, \no \\
l_\mu = v_\mu - a_\mu  &\Gto &  g_{_L} l_\mu g_{_L}^{-1}, \no \\ 
s + ip   &\Gto &  g_{_R} (s+ip)g_{_L}^{-1}, \no \\ 
s - ip   &\Gto &  g_{_L} (s-ip)g_{_R}^{-1},
\label{trasfGglob}  
\eeqa
and we consider the Lagrangian
\beq 
\cL_{QCD}(v,a,s,p) = \cL^{(0)}_{QCD}
+\bar{\psi} \gamma^\mu (v_\mu +a_\mu \gamma_5) \psi 
-\bar{\psi} (s - i p \gamma_5) \psi~. 
\label{lqcdvasp}
\eeq
In this way we reach two interesting results:\cite{Ecker}
\begin{itemize}
\item{} The generating functional 
\beq
e^{i Z(v,a,s,p) } = \int {\cal D}q  
{\cal D}\bar{q}{\cal D}G \ e^{ i  \int {\rm d}^4x
\cL_{QCD}(v,a,s,p) } 
\label{zvasp}
\eeq
is explicitly invariant under chiral transformations, but 
the explicit breaking of $G$  can  directly be obtained  
by calculating the  Green functions, i.e.~the functional derivatives of 
$Z(v,a,s,p)$, at 
\beq
v_\mu=a_\mu=p=0 \qquad\qquad s=M_q=\mbox{\rm diag}(m_u,m_d,m_s).
\eeq
\item{} The global symmetry $G$ can be promoted to a local 
one by modifying the transformation laws of 
$l_\mu$ and $r_\mu$ in
\beqa
r_\mu = v_\mu + a_\mu  &\Gto &  g_{_R} r_\mu g_{_R}^{-1}
+i g_R\partial_\mu g_R^{-1}~, \no \\
l_\mu = v_\mu - a_\mu  &\Gto &  g_{_L} l_\mu g_{_L}^{-1}
+i g_L\partial_\mu g_L^{-1}~.  
\label{trasfGloc}  
\eeqa
Then the gauge fields of electroweak interactions
can be automatically included as external fields by the 
substitution
\beqa
v_\mu &\to& -eQA_\mu - {g\over 2\cos\theta_W }\left[ Q\cos(2\theta_W)
-{1\over6} \right]  Z_\mu                          
- {g \over 2\sqrt{2} }\left( T_+ W^+_\mu +
\mbox{\rm h.c.}\right)~, \no \\ 
a_\mu &\to& + {g\over 2\cos\theta_W } \left[ Q -{1\over6} \right]  Z_\mu
+ {g \over 2\sqrt{2} }\left( T_+ W^+_\mu +
\mbox{\rm h.c.}\right)~, 
\label{AWfields}   
\eeqa
where 
\beq
Q={1\over 3}\left(\ba{ccc} 2 & 0 & 0 \\ 0 & -1 & 0 \\ 0 & 0 & -1 \ea \right)~,
\qquad\qquad
T_+=\left(\ba{ccc} 0 & V_{ud} & V_{us} \\ 0 & 0 & 0 \\ 0 & 0 & 0 \ea \right)~,
\eeq
and $V_{ij}$ denote the elements of the Cabibbo--Kobayashi--Maskawa matrix.

As a consequence, Green functions for processes with external photons, $Z$
or $W$ bosons, can simply be obtained as functional derivatives of
$Z(v,a,s,p)$.
\end{itemize}
The chiral realization of $Z(v,a,s,p)$ at the 
lowest order in the derivative expansion is given by the classical action
obtained from ${\widetilde \cL}_S^{(2)}$, after having included
the external sources  in a chiral invariant way.
Concerning spin-1 sources, this can be achieved by means of the minimal
substitution
\beq
\partial_\mu U \to  D_\mu U = \partial_\mu U -ir_\mu U +i Ul_\mu~
\label{eq:minsub}
\eeq
or, in principle,  introducing new operators written in terms of the tensors
\beqa
F_L^{\mu\nu} &=& \partial^\mu l^\nu - \partial^\nu l^\mu -i\left[
l^\mu, l^\nu \right]~, \no \\                                    
F_R^{\mu\nu} &=& \partial^\mu r^\nu - \partial^\nu r^\mu -i\left[
r^\mu, r^\nu \right]~.
\eeqa       
From Eq.~(\ref{eq:minsub}) it appears convenient 
to assign the same power counting to derivatives of
$u$ and to the sources $a^\mu$ and $v^\mu$, so that 
$D_\mu U$ becomes a homogeneous term of first order 
in the derivative (or chiral) expansion:
\beqa
U       &\sim & \cO(p^0)~,  \no \\
u^\mu,~a^\mu,~v^\mu &\sim & \cO(p^1)~, \no \\
F^{\mu\nu}_{L,R} &\sim & \cO(p^2)~.
\label{count1}
\eeqa
According to this assignment, the Lorentz-invariant 
terms containing the external tensors $F_{L,R}^{\mu\nu}$ 
are at least of $\cO(p^4)$ and do not appear at 
the lowest order.

Regarding spin-0 sources, the most natural power-counting
assignment is given by:\cite{GL1}
\beq
\qquad~\ s,~p    \ ~\sim~\ \cO(p^2)~.
\label{sceltasp}
\eeq
As we shall see in the following, this choice 
is welcome since it implies $M^2_\pi \sim \cO(p^2)$ 
and leads to the Gell-Mann--Okubo mass formula.\cite{Gellmann,Okubo}

We are now able to write down the most general Lagrangian
invariant under $G$, of order $p^2$, which includes
pseudoscalar mesons and external sources. This is
\beq
\cL_S^{(2)} = {F^2 \over 4} 
\la D_\mu U D^\mu U\da + \chi U\da + U\chi\da \ra~,
\label{ls2}
\eeq
where 
\beq
\chi = 2B(s+ip)~.
\label{chi}
\eeq
$\cL_S^{(2)}$ is completely determined by chiral 
symmetry but for the couplings $F$ and $B$, which 
have to be constrained from experimental data. 
These two couplings are in turn related to two fundamental
order parameters of the spontaneous breakdown 
of chiral symmetry: the pion decay constant $F_\pi$, defined by
$\bra{0}| \bar{\psi} \gamma^\mu \gamma_5 \psi \ket{\pi^+(p)} =
i \sqrt{2} F_\pi p^\mu$,
and the quark condensate $\bra{0}|\bar{\psi} \psi \ket{0}$.
Indeed, differentiating with respect to the external sources, we find
\beqa
\bra{0}| \bar{\psi} \gamma^\mu \gamma_5 \psi \ket{\pi^+(p)} &=&
 \bra{0}| {\delta S^{(2)} \over \delta a_\mu }  \ket{\pi^+(p)} 
=  i \sqrt{2}  F  p^\mu~, \label{chiralF}  \\
\bra{0}|\bar{\psi} \psi \ket{0} &=& -
\bra{0}| {\delta S^{(2)} \over \delta s }  \ket{0} 
= -F^2 B~,\label{chiralB}
\eeqa
where $S^{(2)}$ is the classical action:
\be
S^{(2)}=\int d^4x \cL_S^{(2)} \fs
\ee
It is important to stress that relations (\ref{chiralF}--\ref{chiralB}) 
are exactly valid only in the chiral limit: 
in the real case ($m_q\not=0$) they are  
modified by $\cO(m_q^2) \sim \cO(p^4)$ terms.  

The pion decay constant is experimentally  known from the process
$\pi^+\to \mu^+ \nu$: $F_\pi=92.4$ MeV. In principle one could 
determine $F$ also from the kaon decay constant, defined
analogously to $F_\pi$ and measured to be $F_K=114$ MeV. 
The difference between 
$F_\pi$ and  $F_K$ is a typical  $\cO(p^4)$ effect,  
which goes beyond lowest order. However, 
since $\cO(p^4)$ effects are expected to be 
larger in the case of $F_K$, the most natural 
determination of $F$ at $\cO(p^2)$ is provided by $F_\pi$.

Contrary to the decay constants, the quark condensate 
is not directly related to any physical
observable. It is the product $B\times m_q$ that can be constrained by
means of experimental data. This appears in the quadratic terms of
$\cL_S^{(2)}$ and is therefore related to pseudoscalar meson masses:
\beqa
M_{\pi^+}^2 &=& (m_u+m_d)B~, \no\\
M_{K^+}^2 &=& (m_u+m_s)B~,  \no\\
M_{\Ko}^2 &=& (m_d+m_s)B~.  \label{masseP} 
\eeqa
The analogous equation for $M_{\eta_8}^2$ contains no new parameter 
and gives rise to a consistency relation: 
\beq
3M_{\eta_8}^2=4M_{K}^2-M_{\pi}^2~,
\label{GOkuborel}
\eeq
the famous  Gell-Mann--Okubo mass formula,\cite{Gellmann,Okubo}
well satisfied by experimental data under the 
assumption $M_{\eta}=M_{\eta_8}$. The validity of the 
Gell-Mann--Okubo relation provides a significant
a posteriori check that the $\cO(m_q^2)$ corrections to 
Eqs.~(\ref{masseP}) are small and that 
the assignment $s \sim \cO(p^2)$ is consistent.

\subsection{Quark-mass ratios and isospin breaking}

Similarly to the quark condensate, also the absolute value 
of light-quark masses cannot be determined within CHPT. 
Relations (\ref{masseP}) provide, however, stringent
constraints on quark mass ratios.  

In addition to $\cO(m_q^2)$ corrections, relations (\ref{masseP})
can be affected by electromagnetic effects. At leading order 
in the chiral expansion, the latter can only depend on meson charges, 
and we can therefore write
\beqa
M_{\pi^0}^2 &=& (m_u+m_d)B + \cO(m_q^2)  ~, \no\\
M_{\Ko}^2 &=& (m_d+m_s)B   + \cO(m_q^2)  ~, \no\\
M_{\pi^+}^2 &=& (m_u+m_d)B + \alpha\Delta_{\rm e.m.} + \cO(m_q^2, \alpha m_q)  ~, \no\\
M_{K^+}^2 &=& (m_u+m_s)B   + \alpha\Delta_{\rm e.m.} + \cO(m_q^2, \alpha m_q)  ~.   
\label{masseP2} 
\eeqa
Neglecting the small $\cO(m_q^2, \alpha m_q)$ corrections
and using the experimental values of pseudoscalar meson masses, 
one can extract from Eqs.~(\ref{masseP2}) the following
two ratios
\beqa
\frac{m_d - m_u}{m_d + m_u} &=& \frac{  (M_{\Ko} -M_{K^+}) 
 - (M_{\pi^0} - M_{\pi^+}) }{M_\pi^0}
 =  0.22 + 0.07 = 0.29~, \label{eq:mud} \\
\frac{m_s}{m_d + m_u} &=& \frac{(M_{K^+}-M_{\pi^+}) + (M_{\Ko} - M_{\pi^0}) }{M_\pi^0}
 =  25~.
\eeqa
Interestingly, the three light-quark masses turn out to be rather different: 
Eq.~(\ref{eq:mud}) indicates a sizable violation ($\sim 30\%$)
of isospin symmetry and Eq.~(\ref{eq:mud}) shows 
that $SU(3)$ is not at all a good symmetry for quark masses. 
However, it is known that both symmetries,
especially the isospin one, are usually 
well respected by strong interactions. 
For instance, considering the spectrum of 
vector mesons, it is found that isospin works at the (1--2)\%
level [$(M_\omega-M_\rho)/M_\rho \sim 1.5\%$] and  $SU(3)$
at the (10--20)\% level [$(M_{K^*}-M_\rho)/M_\rho \sim 16\%$].

The reason of this behaviour can be traced back to the 
smallness of quark masses with respect to the scale of chiral
symmetry breaking. Indeed, although we cannot have a precise 
information about the absolute value of quark masses, 
by looking at the breaking of isospin and $SU(3)$ 
symmetries in the spectrum of hadrons we can infer 
the following hierarchy: 
\beqa
\Lambda_\chi & \sim & 10^3~ {\rm MeV}~, \no \\
         m_s & \sim & 10^2~ {\rm MeV}~, \no \\
    m_d,~m_u & \sim & 10^1~ {\rm MeV}~.
\eeqa
The masses of all light hadrons, except for the would-be
Goldstone bosons,  remain different from zero in the 
chiral limit and are mainly determined by $\Lambda_\chi$.
Then isospin and $SU(3)$ symmetries are {\em accidental}
consequences of the fact that  
$m_{u,d,s} \ll \Lambda_\chi$.
The observation that  both these symmetries
are well respected in the hadronic world 
is a good a posteriori check of
the basic assumptions of CHPT discussed in 
Section~\ref{sect:chsymm}.

\subsection{Lagrangian at order $p^4$}
\label{sec:p4}
The Lagrangian (\ref{ls2}) is only the first term of an infinite series. At
the moment, in the purely strong sector the Lagrangian is known up to and
including order $p^6$.\cite{BCE} Here we introduce the Lagrangian of order
$p^4$ that was originally written down by Gasser and Leutwyler.\cite{GL1}
In order to derive this Lagrangian, one needs to list all possible
chiral-invariant terms of order $p^4$. It is not strictly necessary, but very
useful, to also reduce this list to a minimum, using all possible trace
identities (which depend on the number of light flavours) and also the
classical equations of motion derived from the Lagrangian (\ref{ls2}), in
all possible ways. The interested reader is referred to the original
articles, \cite{GL1,BCE} for more details on how this is done. Here we
simply write down the Lagrangian in the case of three light flavours:
\bea
\label{L4}
{\cal L}_4 & = & L_1 \langle D_\mu U^\dagger D^\mu U\rangle^2 +
                 L_2 \langle D_\mu U^\dagger D_\nu U\rangle
                     \langle D^\mu U^\dagger D^\nu U\rangle \no \\*
& & + L_3 \langle D_\mu U^\dagger D^\mu U D_\nu U^\dagger D^\nu U\rangle +
    L_4 \langle D_\mu U^\dagger D^\mu U\rangle \langle \chi^\dagger U +
    \chi U^\dagger\rangle  \no \\*
& & +L_5 \langle D_\mu U^\dagger D^\mu U(\chi^\dagger U + U^\dagger \chi)\rangle
    +
    L_6 \langle \chi^\dagger U + \chi U^\dagger \rangle^2 +
    L_7 \langle \chi^\dagger U - \chi U^\dagger \rangle^2  \no \\*
& & + L_8 \langle \chi^\dagger U \chi^\dagger U +
 \chi U^\dagger \chi U^\dagger\rangle
    -i L_9 \langle F_R^{\mu\nu} D_\mu U D_\nu U^\dagger +
      F_L^{\mu\nu} D_\mu U^\dagger D_\nu U \rangle \no \\*
& & + L_{10} \langle U^\dagger F_R^{\mu\nu} U F_{L\mu\nu}\rangle +
    L_{11} \langle F_{R\mu\nu} F_R^{\mu\nu} + F_{L\mu\nu} F_L^{\mu\nu}\rangle +
    L_{12} \langle \chi^\dagger \chi \rangle \co
\eea
where, for completeness, we have also written down the so-called ``contact
terms'', those multiplying the constants $L_{11}$ and $L_{12}$, which
contain only external fields. We should add that at this order also the
Lagrangian of Wess, Zumino and Witten\cite{WZW} enters. This describes the
effects due to the axial anomaly, which are therefore parity-violating. 
In the following we will not consider such effects.


\section{One-loop graphs: renormalization}

In the following two sections we will introduce the subject of loop
calculations in CHPT. Technically, these loop calculations need no
special introduction: any graduate student who has already made loop
calculations (in QED for example) should be able to perform them also in CHPT
also. On the other hand, at a conceptual level, he/she may have doubts about
the meaning of these loop calculations: how should one interpret the loops
of pions, which contain contributions of pions of any virtuality, if one
knows that at high energy scales these degrees of freedom are no
longer relevant? The main aim of these two lectures is to answer these kinds of
questions, and to illustrate the physical meaning of the loop contributions. 

\subsection{The scalar form factor of the pion}

To illustrate the basic concepts of loop calculations in CHPT we will focus
on one specific example: the scalar form factor of the pion, defined as
\be
\langle \pi^i(p_1) \pi^j(p_2)|\hat m(\bar u u + \bar d d)|0\rangle =:
\delta^{ij} \Gamma(t) \co \; \; \; \; t=(p_1+p_2)^2~,
\ee
where $\hat m= (m_s+m_d)/2$.
This matrix element is relevant to the decay $h \to \pi \pi$, which
would have been the main decay mode for a
light Higgs (of course this scenario is now experimentally excluded).
The tree-level calculation of this matrix element is simple: in
the Lagrangian (\ref{ls2}) the coupling between the scalar source and two
pions, which can be read from Eq.~(\ref{chi}), does not involve derivatives
and leads to
\be
\Gamma(t)=2\hat m B = M_\pi^2+\cO(p^4) \fs
\ee
This result, which we worked out from the Lagrangian, is actually a
consequence of a general theorem, due to Feynman and Hellman.\cite{FH} 
This states that the expectation value of the perturbation in an
eigenstate of the total Hamiltonian determines the derivative of the energy
level with respect to the strength of the perturbation:
\be
\hat m {\partial M_\pi^2 \over \partial \hat m} = \langle \pi |\hat m \bar
q q | \pi \rangle = \Gamma (0) \fs
\ee
The value of the form factor at zero momentum transfer is fixed by this
theorem, and a simple power counting implies that at leading order the
scalar form factor is a constant -- at order $p^2$ the theorem completely
fixes the form factor. On the other hand no general principle forbids a
dependence of the form factor on $t$ (to the contrary, they imply it, as we
will see), and to generate this we necessarily have to go beyond leading
order. 

\subsection{$SU(2)$ Lagrangian at order $p^4$}
Before starting the loop calculation, let us have a look at what happens to
the form factor once we include tree-level contributions from higher-order
terms in the Lagrangian. The Lagrangian at order $p^4$ has been discussed 
in Sect. \ref{sec:p4} in the case of three light flavours. For the case at
hand, the role of kaons and etas is marginal, if we stick to the very
low-energy region. It is more convenient to use a simpler Lagrangian, by
expanding around $m_u=m_d=0$ and keeping $m_s$ at its physical value. In
this case the chiral symmetry is $SU(2)_L\times SU(2)_R$. The Lagrangian
of order $p^2$ remains unchanged -- the only change is that the field $U$,
and its logarithm $\Phi$ are now $2\times 2$ matrices. At order $p^4$ the
Lagrangian is simpler, because we can use more trace identities to reduce
the number of independent terms. For two light flavours this is a sum of
seven terms\footnote{The ellipsis is for the so-called contact terms,
  i.e. those that depend only on the external sources.}
\be
\cL_S^{(4)} = \sum_{i=1}^7 l_i P_i + \ldots \co
\label{ls4}
\ee
where
\bea
&&P_1=\frac{1}{4} \langle u_\mu u^\mu \rangle^2 \co ~~~
P_2=\frac{1}{4}\langle u_\mu u_\nu \rangle \langle u^\mu u^\nu
\rangle \co \nn
&& P_3= \frac{1}{16} \langle \chi_+ \rangle^2 \co ~~~
P_4= \frac{i}{4} \langle u_\mu \chi_-^\mu \rangle \co ~~~
P_5=-\frac{1}{2}\langle f_{- \mu \nu} f_-^{\mu \nu} \rangle \co \nn
&&P_6=\frac{1}{4} \langle [u_\mu,u_\nu]f_+^{\mu \nu} \rangle \co ~~~ 
P_7= -\frac{1}{16} \langle \chi_- \rangle^2 \co
\eea 
and we have used the compact notation:
\bea
\chi_\pm &=&  u^\dg \chi u^\dg \pm u \chi^\dg u \nn  
\chi_\pm^\mu &=& u^\dg D^\mu \chi u^\dg \pm u D^\mu \chi^\dg u \nn
f_{\pm }^{\mu\nu} &=& u F_{L}^{\mu\nu} u^\dg \pm u^\dg F_{R}^{\mu\nu} u \fs
\eea
Only two of these seven terms contribute to the scalar form factor:
the terms proportional to $l_3$ and $l_4$. Their contribution reads:
\be
\Gamma_{[l_3,l_4]}(t) = {M^2 \over F^2} \left[ -4 M^2 l_3 + t l_4\right]\; ;
\label{Gl3l4}
\ee
the calculation is recommended as an easy exercise, as is the
calculation of the contribution of $l_3$ to the pion mass. Once these two
calculations are completed, one can then check that the Feynman--Hellman
theorem is respected also in this case.

The result in Eq. (\ref{Gl3l4}), a tree-level calculation with the
next-to-leading order Lagrangian, is merely a statement about how the
symmetry constrains this particular matrix element: at this order the scalar
form factor can have at most a term linear in $t$. No symmetry relation
exists between the constant term and the coefficient of the linear term,
hence we have two different constants. The constant term is related to the
derivative of the pion mass with respect to the strength of the
symmetry-breaking term in the Lagrangian, whereas the coefficient of the
linear term is related to the correction to the pion decay constant (again,
it is a very good exercise to calculate the latter with the Lagrangian in
Eq. (\ref{ls4})).

\subsection{Loop graphs}
\begin{figure}[t]
    \begin{center}
       \setlength{\unitlength}{1truecm}
       \begin{picture}(6.0,3.)
       \epsfxsize 6.  true cm
       \epsffile{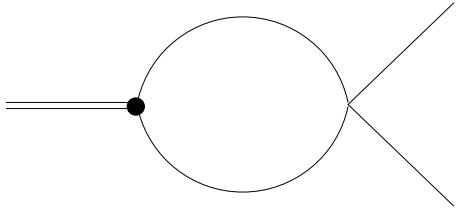}
       \end{picture}
    \end{center}
    \caption{\label{fig:loop1} One-loop diagram for the scalar form factor
      of the pion. The double line stands for the scalar source, whereas
      single lines for pions.}
\end{figure}
If we neglect the tadpole graphs and those renormalizing the external legs,
there is basically only one graph for this process, the one shown in
Fig. \ref{fig:loop1}. Its structure is 
\bea
\int{ d^4l \over (2 \pi)^4} {\{ M^2, p^2,p\!\cdot\! l, l^2\} \over
  (l^2-M^2)((p-l)^2 -M^2)}&\Longrightarrow& x \int { d^4l \over (2 \pi)^4} {1 
    \over (l^2\!-\!M^2)} \nn
&+&  y \int { d^4l \over (2 \pi)^4} { 1 \over
  (l^2\!-\!M^2)((p\!-\!l)^2\!-\!M^2) }  \nn
&\equiv& x T(M^2)+  yJ(t) \co
\label{li}
\eea
where $p=p_1+p_2$. We have indicated, in the first integral, all the
terms that can appear in the numerator, and, after the arrow, the two
possible structures to which the various terms can be reduced.
The momenta and masses in the numerator come from the four-pion vertex on
the right-hand side of the diagram. The power counting for this integral
shows that it represents a correction of order $p^2$ to the leading-order
term (the integration measure, which is of order $p^4$, is compensated by
the two propagators). This is true in general: one-loop graphs constructed
from the Lagrangian (\ref{ls2}) are always a correction of
next-to-leading order in the chiral expansion. For example, it is easy to
see that no matter how many vertices we add on the internal lines in the
graph in Fig. \ref{fig:loop1}, the graph remains of order $p^2$: if we add a
vertex of order $p^2$ we also get an extra propagator, which compensates it.
The general theorem was first proved by Weinberg.\cite{Weinberg79} We will
come back to this point later. 

If we expand the tadpole integral $T(M^2)$ and the
loop integral $J(t)$ in a Taylor series in their respective arguments:
\bea
T(M^2)&=& a+bM^2+\bar T(M^2) \co \nn
J(t)&=& J(0)+\bar J(t) \co
\label{loop_int}
\eea
only the first terms in the expansion are divergent, whereas both
$\bar T(M^2)$ and $\bar J(t)$ are finite -- this is easily seen by taking a
sufficient number of derivatives on the loop integrals in Eq. (\ref{li}).
It is left as an exercise to the reader to show that the one-loop diagrams
that we have neglected can only produce terms like
$T(M^2)$. Therefore the contribution of the loop diagrams to the scalar
form factor has the following structure:
\be
\Gamma(t) \sim
{M^2 \over F^2} \left[ x_1 b M^2 + x_2 t J(0)  + x_1 \bar T(M^2) + x_2 \bar
  J(t) \right]  \fs
\ee
The divergent part of the loop diagrams has exactly the same structure as
the counterterm contribution calculated above: to remove it we
simply need to define the counterterms properly (in this case the constants
$l_3$ and $l_4$). 

The principles stated in the first lecture that guided the construction of
the effective Lagrangian only appealed to symmetry arguments, and therefore
allowed for an infinite number of terms. Once this is accepted, there is no
problem of principle in carrying through the renormalization program: as
anticipated, the difference between renormalizable and non-renormalizable
theories is, in a sense, a technical detail.

\subsection{Chiral logarithms}
If we expand the form factor in a Taylor series in $t$, we can write it in
the following form:
\be
\Gamma(t) = \Gamma(0) \left[1+ {1 \over 6} \la r^2 \ra^\pi_S t+O(t^2)
\right] \fs
\ee
The coefficient of the linear term, properly normalized, is called the
scalar radius of the pion. Its size is a way to represent the spatial
extension of the pion when probed with a scalar source. We have stated
above that the coupling constant that appears in this quantity, $l_4$, also
determines the first correction of the pion decay constant around the
chiral limit. There is another piece of information on the scalar radius, which we
can already gather from the simple sketch of the loop calculation given
above:
\be
\langle r^2 \rangle^\pi_S \sim J(0)= \int {d^4l \over
  (2\pi)^4} {1 \over (l^2-M^2)^2} \sim \ln {M^2 \over \Lambda^2} \co
\ee
namely that the scalar radius contains an infrared divergence. In the
chiral limit this quantity diverges. This divergence should not be removed
and does not represent a problem because it has a physical meaning, in the
sense that when the pion becomes massless the cloud of pions surrounding any
hadron (and therefore also the pion itself) extends to an infinite
range. A quantity that measures the spatial extension of this
cloud should indeed become infinite in the chiral limit.
Notice that the scalar form factor is finite and remains finite also in the
chiral limit -- it is only the first derivative in $t$, calculated at $t=0$
that diverges when the pion mass goes to zero.

These infrared divergences are present everywhere in pion physics, and in
many cases they are among the most important physical effects (less so in
the case of kaons). Their relevance was first pointed out by Li and
Pagels.\cite{LiPagels} The effective Lagrangian method provides a
systematic way to calculate these effects. As we have seen in the above 
example, they arise from the infrared region in the loop integrals --
precisely the region where we should fully trust the vertices of our
effective Lagrangian.

One may be less at ease with the ultraviolet region of the loop integrals:
there indeed has no justification the use of the effective
Lagrangian. On the other hand, through the process of renormalization, that
part of the integrals is completely removed and substituted with unknown
constants, the counterterms. As is sometimes said, these parametrize our
ignorance of the physics that lies above the range of applicability of the
effective Lagrangian. In fact, we are not completely ignorant about the
physics of strong interactions in the GeV range and above, and as we will
see in the following lecture, it is not difficult to get a good estimate of
the numerical value of the counterterms.

Until now the only part of the loop integrals that we have not analysed
is the finite, analytically non-trivial part of the loop integral, the
function $\bar J (t)$. This will be  treated in full detail in the
following lecture. Before doing that, however, we want to consider the
problem of renormalization from a more general point of view.

\subsection{Renormalization and chiral symmetry}
The reader who has done the exercise of calculating the divergent part of
the scalar form factor to one loop will have seen that, to renormalize it,
the two counterterms $l_3$ and $l_4$ have to be defined in the following way
\be
l_3 = l_3^r(\mu) - {1 \over 2} \lambda \co ~~~ l_4 = l_4^r(\mu) + {2}
\lambda \co 
\ee
where $\lambda$ is divergent and dependent on the regularization method
(in dimensional regularization, for example, it is defined by
$\lambda=(c\mu)^{d-4}/(d-4)$, with $c$ an arbitrary constant that defines
the regularization scheme).

If we look at the definition of the operators in front of $l_3$ and $l_4$
we see that (as usual) they contain an infinite number of pion fields.
For example they both contain a term with one scalar source and 6 pion
fields. Does it mean that if we had calculated that matrix element to one
loop we would have found the same divergent part for $l_3$ and $l_4$? Or,
to put it differently, that once we calculate the divergent part of the
scalar form factor then we know the divergent part of all other matrix
elements with a higher number of pion fields\footnote{Strictly speaking,
  these questions make full sense only if we suppose that these two are
  the only counterterms at order $p^4$.}?  

The answer is yes. Chiral symmetry puts a strong constraint on the
divergences: they have to be chiral-invariant terms. This conclusion is
originally due to Weinberg,\cite{Weinberg79} on the basis of a highly
plausible, but still heuristic argument. It is now put on a solid basis by
the work of Leutwyler:\cite{Leu96} he proved that to calculate hadronic
Green functions with an effective Lagrangian, such that they respect the
Ward identities implied by the chiral symmetry, one necessarily has to
start from a chiral-invariant effective Lagrangian.  The non-trivial part in
this statement is that it takes into account also quantum effects:
anomalous symmetries show that it is not always true that a symmetry that
exists at the classical level survives the quantum corrections -- or
vice versa, that to have a symmetrical quantum theory one necessarily has to
start from a symmetrical classical Lagrangian.

The theorem applies also to the divergent part of the quantum corrections:
they have to be chiral-invariant. A general theorem of quantum field theory
states that the divergent part of a loop graph is a polynomial in the
external masses and momenta. These two theorems lead to the conclusion
that the divergences, order by order, can be reabsorbed by the
chiral-invariant counterterms.

While these general theorems are very important, performing a direct
calculation of the divergences and the subsequent renormalization in an
explicitly chiral-invariant form, is probably much more instructive.
The rest of this lecture will be dedicated to an illustration of the
tools necessary for this calculation and to a sketch of the
procedure.

\subsection{Background field method and heat kernel}
\begin{figure}[t]
    \begin{center}
       \setlength{\unitlength}{1truecm}
       \begin{picture}(8.0,3.)
       \epsfxsize 8.  true cm
       \epsffile{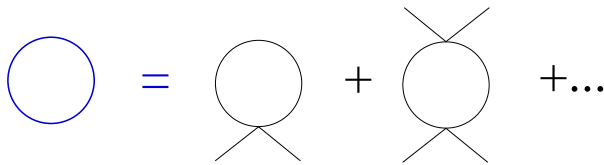}
       \end{picture}
    \end{center}
    \caption{\label{fig:oneloopsum} Graphical representation of the one-loop 
      contributions to the generating functional. The logarithm of the
      differential operator, represented as the loop without external legs,
      is the sum of all diagrams with any number of external legs.}
\end{figure}

To introduce these techniques, it is useful to consider a simpler
theory than CHPT: a purely scalar $O(N)$ $\phi^4$ theory. 
Its Lagrangian is the following:
\begin{eqnarray}
{\cal L}&=&{\cal L}_0+\sum_{n=1}^\infty \hbar^n {\cal L}_n \co \nn
{\cal L}_0&=& {1 \over 2} \left(\partial_\mu \phi^i \partial^\mu \phi^i-
  M^2 \phi^i \phi^i\right)-{g \over 4} \left( \phi^i \phi^i\right)^2 -\phi^i
f^i \co
\end{eqnarray}
where $\cL_0$ is the classical Lagrangian and $\cL_i$, $i\ge 1$, the series
of counterterms needed to renormalize the theory, and a summation over
repeated indices is implied. So as to make the situation
completely analogous to CHPT, where we have the external sources $v, \; \;
a, \; \; s$ and $p$, we have also introduced
external sources $f^i$ coupled to the fields $\phi^i$. 
A path integral constructed with this Lagrangian is a
function of the external sources $f^i$. By taking the appropriate
functional derivatives with respect to the sources we can extract from the
path integral all possible Green functions of the fields $\phi^i$. The
logarithm of the path integral is usually called the generating functional
$Z\{f\}$:
\begin{eqnarray}
e^{iZ\{f\}/\hbar}&=&N \int [d\phi]e^{iS/\hbar}\co \qquad \quad
\qquad \qquad S=\int dx
{\cal L} \co \nn
Z&=&Z_0+\hbar Z_1+\hbar^2 Z_2+O(\hbar^3)\co \qquad Z\{0\}=0  \fs
\end{eqnarray}
To calculate the classical part of the generating functional $Z_0$ and its
quantum corrections $Z_i$, it is useful to expand the field $\phi$ around
the solution of the classical equations of motion, $\bar \phi$:
\be
{\delta S_0 \over \delta \phi^i}=0 \qquad \Rightarrow \qquad (M^2+\Box )
\bar{\phi}^i+g \bar \phi^2 \bar \phi^i+f^i=0 \fs
\ee
We can then shift the integration variable:
\be
\phi^i=\bar \phi^i+\xi^i \co \qquad [d\phi]=[d\xi]\co \qquad
\xi \sim O(\hbar^{1/2}) \co
\ee
and get, for the path integral:
\be
e^{iZ\{f\}/\hbar}=Ne^{i\bar S/\hbar} \int [d\xi] \exp \left\{{i\over \hbar}
  \int dx \left[ {1 \over 2} \xi^i D_{ij} \xi^j+O(\xi^3) \right] \right\}
\co \label{GFexp}
\ee
where
\begin{eqnarray}
D_{ij}&=&-\Box \delta_{ij}+\sigma_{ij} \co \nn
\sigma_{ij}&=&-\left(M^2 + g \bar \phi^2 \right) \delta_{ij}-2g \bar \phi^i
\bar \phi^j \fs
\end{eqnarray}
The separation of the field into its classical part $\bar \phi$ and its
quantum fluctuations $\xi$ makes the calculation of the Taylor series of
$Z$ in $\hbar$ more transparent. Indeed one immediately sees that the
classical action $S_n$ of the Lagrangian $\cL_n$, evaluated at the solution
of the classical equation of motion contributes only to the term $Z_n$ of
the generating functional. The path integral over the quantum fluctuations
yields higher-order quantum corrections. For example $Z_1$ receives a
contribution from the integral over the quadratic term in $\xi$~, which was
explicitly given in Eq. (\ref{GFexp}). The path integral of the
exponential of a quadratic term is known, and can be given in closed form
-- it is the determinant of the differential operator $D_{ij}$: 
\be
Z_1=\int dx \left[ {i \over 2} \ln \left(D_{ij} {D_0}^{-1}_{ij} \right) +
  {\cal L}_1 \right] \co
\label{z1}
\ee
where $D_0=D_{|f=0} = -(\Box+M^2) \delta_{ij}$. To acquire familiarity with
the formal expression (\ref{z1}) it is useful to expand the operator $D$
around $f=0$: $D_{ij} = {D_0}_{ij}+\eta_{ij}$, and correspondingly expand
the logarithm around 1:
\be
\ln\left(D_{ij} {D_0}^{-1}_{ij} \right) = \eta_{ij}{D_0}^{-1}_{ij}-
{1 \over 2} \eta_{ij}{D_0}^{-1}_{jk} \eta_{kl}{D_0}^{-1}_{li} + \ldots \fs
\ee
It should now be clear that this logarithm simply represents the sum of all
one-loop diagrams with any number of four-$\phi$ vertex insertions,
as shown in Fig. \ref{fig:oneloopsum}.
The reader with some experience of loop calculation will have immediately
realized that only the first two terms in the expansion of the logarithm
generate divergences: all the others are finite loop integrals. Indeed the
divergent part of $Z_1$ can be calculated by working out two simple loop
diagrams.

The situation would not be as simple in the case of CHPT: there, the
presence of derivative couplings complicates the situation
significantly. The calculation can nonetheless be done in close form with
the help of the heat kernel method. Without giving many details let us see
how it works in the case of the $O(N)$ theory.

A differential operator of the form:
\be
D^2=-d^2+\sigma \co \qquad d_\mu=\partial_\mu+\gamma_\mu  \co
\ee
where $\sigma$ and $\gamma$ are matrices in general,
is called an elliptic differential operator, and has the same form as the
differential operator that appears in the equations describing the
diffusion of the heat (which explains the origin of the name). The
divergent part of the logarithm of such operators is known in
terms of the operators $\sigma$ and $\gamma_{\mu \nu}$, without knowing
their explicit form. It is given by:
\be
\int dx \ln \left(D_{ij} {D_0}^{-1}_{ij} \right)={i \over (4
  \pi)^2(d-4)} \int dx \left[{1 \over 6} \gamma_{\mu \nu} \gamma^{\mu \nu}+
  \sigma^2 \right] + \ldots \fs
\ee
Details on how this result is obtained can be found in Ref.\cite{JO}.
In the case of the $O(N)$ theory that we are considering, $\gamma_\mu=0$
and we simply have to calculate the square of $\sigma$:
\be
\sigma^2= 2 (N+2)g M^2 \phi^2+ (N+8) g^2 \phi^4\; .
\ee
The calculation of the divergences is complete. The reader is urged to
verify this result by an explicit calculation of the two divergent diagrams
in Fig. \ref{fig:oneloopsum}.

\subsection{Calculation of the one-loop divergences in CHPT}
With the tools introduced in the preceding section we have reduced the
problem of the calculation of one-loop divergences to a simple algebraic
exercise. Even in the case of CHPT. The only step that requires some care
is the choice of the field $\xi$, representing the fluctuations around the
classical solution. Indeed this can be done in many different ways, but
only a few of them will simplify and make  the intermediate
steps of the calculation transparent. One of the most convenient choices is the
following:\cite{GL1}
\bea
U&=&e^{i\phi/F} \co \qquad \qquad \phi=\bar \phi +\xi \co \nn
U&=&\bar u e^{i\xi/F} \bar u \fs
\eea
%
As we have seen in the first lecture, 
the transformation properties of the exponential of the $\xi$
field are particularly well suited to check the chiral invariance 
properties during the various steps of the calculation.

The expansion of $S_2=\int d^4x \cL_S^{(2)}$ around the classical solution
reads 
\bea
\int dx \cL_S^{(2)} &=& \int dx {F^2 \over 4} \langle u_\mu u^\mu + \chi_+
\rangle \nn
&=& \int dx \bar \cL_S^{(2)} +\int dx {1 \over 2} \xi^i \Delta_{ij} \xi^j
+O(\xi^3) \co
\eea
where
\bea
\Delta&=&-d^2+\sigma \co \nn
d_{\mu,\;ij}&=&\partial_\mu \delta_{ij}+\gamma_{\mu,\;ij} \co \nn
\gamma_{\mu,\;ij}&=&-{1 \over 2} \langle [\lambda_i,\lambda_j] \Gamma_\mu
\rangle \co \nn
\Gamma_\mu &=& {1 \over 2} \left\{u^\dagger (\partial_\mu -i r_\mu)u+
  u(\partial_\mu-il_\mu) u^\dagger \right\} \co \nn
\sigma_{ij} &=& -{1 \over 8} \langle [u_\mu,\lambda_i][\lambda_j,u^\mu]+ \{
\lambda_i,\lambda_j \} \chi_+ \rangle \fs
\eea

It is now a simple algebraic exercise to calculate the divergent part of
the generating functional to one loop ($\epsilon=d-4$):
\bea
\label{logdet}
-\frac{i}{2}\log \, \mbox{det}\, \Delta &=& \frac{1}{(4 \pi)^2} 
\frac{1}{\epsilon}\int d^dx \left\{ \frac{N_f}{96} \langle 
\left([u_\mu, u_\nu]-2i f_{+ \mu \nu}\right)^2 \rangle +
\frac{N_f}{16}  \langle \left(u_\mu u^\mu + \chi_+ \right)^2 \rangle
\right. \no \\*
&& \; \; \; \; \; \; \; \; \; \; \; \; \; \; \; \; \; \; \; 
+\frac{1}{16} \langle u_\mu u^\mu + \chi_+ \rangle^2
+\frac{1}{8} \langle u_\mu u_\nu \rangle^2 \nn
&& \; \; \; \; \; \; \; \; \; \; \; \; \; \; \; \; \; \; \; \left.
-\frac{1}{4 N_f} \langle \chi_+^2 \rangle 
+\frac{1}{8 N_f^2}\langle \chi_+ \rangle^2
\right\} + \ldots \fs
\eea
This expression is valid for a generic number of light flavours $N_f$, and
is explicitly chiral-invariant, as it should be.
To obtain the result in the interesting physical cases $N_f=2,3$ is not
only necessary to trivially substitute $N_f$ with the numerical value of
interest, but also to reduce all the chiral-invariant terms appearing in
Eq. (\ref{logdet}) to a minimal set. This step involves the use of trace
identities and of the equations of motion. Without giving the details of
this final step, let us conclude this lecture by presenting the results in
the case of $N_f=2$. If we define the counterterms as
\be
l_i=l_i^r(\mu)+\gamma_i \lambda(\mu) \co
\ee
their coefficients for the divergent part are:
\bea
 \gamma_1=\frac{1}{3}, &\displaystyle \gamma_2 = \frac{2}{3}, & \gamma_3
=-\frac{1}{2}, \nn
\gamma_4=2, &\displaystyle \gamma_5=-\frac{1}{6}, & \gamma_6 =
-\frac{1}{3}, \; \; \gamma_7=0 \fs
\eea

\section{One-loop graphs: analyticity and unitarity}
According to the Watson theorem,\cite{Watson} above threshold but
below the inelasticity threshold, the phase of the 
scalar form factor is equal to the $S$-wave 
$\pi \pi$ phase shift with isospin $I=0$,
$\delta_0^0$. As is well known, this theorem is a consequence of unitarity:
\bea
\im \Gab(t) &=& \sigma(t) \Gab(t) {t_0^0}^*(t) =
\Gab(t) e^{-i\delta_0^0} \sin \delta_0^0 \nn
&&\\
&=& |\Gab(t)| \sin \delta_0^0 \co
\label{unitarity}
\eea
where $\sigma(t)=[1-4M_\pi^2/t]^{1/2}$ and $\Gab(t)=\Gamma(t)/\Gamma(0)$,
and $t_0^0$ is the $I=0$ $S$-wave of $\pi \pi$ scattering.
The unitarity relation (\ref{unitarity}) shows that the leading-order
expression $\Gab(t)=1$ cannot be the whole story: if we want an
accurate description of the form factor away from $t=0$, we need to include
higher orders and, in particular, loops -- imaginary parts can only be
generated by loop graphs.

Notice that since at leading order $\Gab(t)$ is $\cO(1)$, and the phase
$\delta_0^0$ is $\cO(p^2)$, the $\cO(p^2)$ imaginary part (which is a
next-to-leading order correction) is completely fixed by leading-order
quantities: 
\be 
\im \Gab^{(2)}(t) = {\delta_0^0}^{(2)}(t)=\sigma(t) {2t-M_\pi^2
\over 32 \pi F_\pi^2} \fs 
\label{1l_unitarity}
\ee
The use of the effective Lagrangian method to calculate the form factor
guarantees that this relation is satisfied. The complete expression for the
one-loop scalar form factor reads as follows:
\be
\Gab(t)=1+{t \over 16 \pi^2 F_\pi^2} (\bar l_4-1)+{2t-M_\pi^2 \over
  2 F_\pi^2} \bar J(t) \co
\label{Goneloop}
\ee
where $\bar J(t)$ is the subtracted one-loop integral (\ref{loop_int}). Its
explicit expression reads:
\be
\bar J(t) = {1 \over 16 \pi^2}\left[ \sigma(t) \ln {\sigma(t)-1 \over
    \sigma(t)+1} +2 \right] \fs
\ee
The reader can now easily verify that the imaginary part of the form factor
at this order indeed satisfies (\ref{1l_unitarity}).

\subsection{Dispersion relation for the scalar form factor}
An analytic function must be real on the real axis: the scalar form factor
is non-analytic from threshold ($4 M_\pi^2$) up to infinity. On the basis
of very general arguments, which mainly use the causality principle, one
can prove that as a function of $t$, the scalar form factor must be
analytic everywhere else (see Ref.\cite{WeinQFT} for a general discussion
of the analyticity properties of amplitudes and Green functions). 
Since it is analytic everywhere else, the non-analyticity of the form factor
on the real axis can be further characterized, and described as a 
discontinuity:
\be
\Gab(t+i\varepsilon) = \Gab^*(t-i\varepsilon)  = |\Gab(t)|
e^{i\delta_0^0(t)} \fs
\ee

Given these analytic properties, we can write the following dispersion
relation:
\be
\Gab(t)= 1+b t +{t^2 \over \pi} \int_{4M_\pi^2}^{\infty}
  {dt' \over t'^2} 
  {\im \Gab(t') \over t'-t}  \co
\label{disprel}
\ee
where, for later convenience, we have subtracted the dispersive
integral twice
-- we will come back to the issue of how many subtractions are
necessary for the dispersive integral to converge. The dispersion relation
shows that, if we know the subtraction constants (in this case only one,
$b$) and the imaginary part on the real axis, we can reconstruct the scalar
form factor everywhere on the complex plane.

It is no surprise that any perturbative calculation in a quantum field
theory produces amplitudes and Green functions with the correct analytic
properties. Using an effective field theory makes no difference: the form
factor calculated to one loop in CHPT must have the correct analytic
properties, and must satisfy (at the perturbative level) the dispersion
relation (\ref{disprel}). To convince ourselves that this is actually the case,
let us first apply the chiral counting to the dispersion relation:
\bea
\Gab^{(0)}&=&1 \nn
b&\sim&\cO(1)\nn
\im \Gab(t')&=&\cO(p^2) \fs
\eea
As we have seen the $\cO(p^2)$ imaginary part is fully fixed by leading-order
quantities, (\ref{1l_unitarity}), and apart from an unconstrained
polynomial term, the real part must be given by the dispersive integral
over this known imaginary part. We leave it as an exercise to prove that
this is true. For this it is useful to know that:
\be
\bar J(t)={t \over 16 \pi^2} \int_{4 M_\pi^2}^{\infty} {dt' \over t'}
{\sigma(t') \over t'-t} \fs
\ee

\subsection{High-energy contributions to the dispersive integrals}
In the previous section we showed that the renormalization procedure
removes the contributions to the loop integrals where the momentum squared
of the pions is large. This was reassuring because we cannot hope that our
effective Lagrangian describes highly virtual pions well.
In the present section we are dealing with the contribution to the loop
integrals from real pions: the dispersive integrals.
As we have seen above, these extend all the way up to infinity, as required
by analyticity. In the perturbative expansion that we are 
considering, the imaginary part of the form factor, which is in the
integrand, is evaluated only to leading non-trivial order. This description
of the imaginary part can be reasonably accurate only in the low energy
region: still, in the integral, it is used all the way up to infinity.
How can we trust the dispersive integral?

In fact, we don't. At least not for the contribution that comes from the
high-energy region. Suppose we decide to remove the part of the dispersive
integral from $\Lambda=1$ GeV to infinity. We should then subtract from the
form factor a term like: 
\be 
{t^2 \over \pi} \int_{\Lambda^2}^\infty
{{\delta_0^0}^{(2)}(t') \over {t'}^2(t'-t) } = {t^2 \over \pi}
\int_{\Lambda^2}^\infty {{\delta_0^0}^{(2)}(t') \over {t'}^3}
\left(1+{t\over t'} +O(t^2) \right) \fs
\label{cutoff}
\ee
The Taylor expansion inside the integral can be safely performed because
the CHPT calculation of the form factor is valid only for $t \ll
\Lambda^2$. In the chiral language, this part of the dispersive integral
can be represented as a polynomial series starting at order $p^4$, i.e.~at
an order which is beyond the accuracy at which we are presently working.
This shows that worrying about these contributions to the dispersive
integrals is futile: the only sensible way to improve the evaluation of the
dispersive integral is to go one order higher in the chiral expansion. This
would automatically give a representation of the form factor that contains
the dispersive integral over the imaginary part correct up to order $p^4$.

For those who are interested in the numerics, the first term in the Taylor
expansion of the integral (\ref{cutoff}) is equal to:
\be
{t^2 \over \pi} \int_{\Lambda^2}^\infty {{\delta_0^0}^{(2)}(t') \over
  {t'}^3} = 0.7 \left[t(\gev^2)^2\right] \co
\ee
which means, for $t=(0.5 \gev)^2$ (which is about the upper limit of
validity of the chiral expansion), a 4\% correction to the leading-order
result. Also numerically everything is well under control.

\subsection{Estimate of the low-energy constants}
The chiral representation to next-to-leading order satisfies a dispersion
relation with two subtractions. The number of subtractions in this case is
dictated by the behaviour at infinity of the leading-order chiral phase
$\delta_0^0$. The latter, however, has nothing to do with the physical
reality: it can be shown that, with form factors, one subtraction is already
sufficient for the dispersive integral to converge. To discuss the physical
implications of this, it is useful to consider the vector form factor of
the pion, which is defined as follows:
\be 
\langle \pi^i(p_1)| \bar q \tau^k \gamma_\mu q|\pi^j(p_2) \rangle = i
\epsilon^{ikj} F_V(t) \co \qquad q =\left(\begin{array}{c} u \\ d
  \end{array} \right) \fs
\label{vector}
\ee
It satisfies the following dispersion relation:
\be
F_V(t)=1+{t \over \pi} \int_{4M_\pi^2}^{\infty} {dt' \over t'} {\im F_V(t')
  \over t'-t} \fs
\label{vector_disp}
\ee
This dispersion relation implies that the charge radius (the analogue of
the scalar radius, for the vector form factor) is given by a sum rule:
\be
{1 \over 6} \langle r^2 \rangle^\pi_V={1 \over \pi}\int_{4
  M_\pi^2}^{\infty} dt' {\im F_V(t') \over t'^2} \fs
\label{rvsumrule}
\ee
If one knows the imaginary part of the vector form factor one can calculate
the charge radius.

Within the chiral representation the sum rule (\ref{rvsumrule}) does not
make sense as it stands, because the integral on the right-hand side does
not converge. In CHPT, the charge radius is given in terms of one of the
low-energy constants:
\be
{1 \over 6} \langle r^2 \rangle^\pi_V=
-{1 \over F_\pi^2} \left[
  l_6^r(\mu) - {1 \over 96 
    \pi^2}\left( \ln{M_\pi^2 \over \mu^2} + 1 \right) +
  O(M_\pi^2) \right] \co
\ee
and, in this language, the sum rule (\ref{rvsumrule}) can be read as a way
to determine $l_6^r(\mu)$. Notice that the fact that this constant appears
at all in the chiral representation is a consequence of the non-convergence
of the chiral dispersive integral: that integral would receive a sizeable
contribution from the high-energy region, and there the chiral
representation obviously fails, not even allowing the
calculation to be made.

Having the sum rule at hand, we can attempt to estimate this particular
low-energy constant, from an estimate of the behaviour of the imaginary
part of the form factor. What do we know about it? Actually, this has been
measured in various ways, and a lot of experimental information is
available on this quantity. But rather than collecting all the available
information to make an accurate evaluation of the sum rule, we want to make
a simple, but instructive exercise. Since in this channel (two pions in the
$P$ wave and isospin $I=1$) the $\rho$ resonance contributes, we can try to
give a rough estimate of its contribution to the sum rule. For this it is
sufficient to use the narrow-width approximation:
\be 
\im F_V(t) \sim \pi \delta(t-M_\rho^2)  \co
\label{nawidth}
\ee
which, when inserted in the integral, gives
\be
{1 \over 6} \langle r^2 \rangle^\pi_V = {f F_\rho \over F_\pi^2} {1 \over
  M_\rho^2} \co
\ee
where the two coupling constants
\be
F_\rho=144 \mev \co \qquad f =69 \mev \ee
determine the strength of the coupling of the $\rho$ resonance to the
external vector field and to the pions.

\begin{table}[t]
\begin{center}
\begin{tabular}{|c|r||cllllc|r|}
\hline i & $10^3 L_i^r(M_\rho)$ && $V$ & $A\,$ & $\,S$ &
      $S_1$ & $\eta_1$ & Total
\\ \hline 1 & $0.53\pm0.25$ &\hspace{3mm} &
      $0.6$ & $0$ & $\hspace{-3.3mm}-0.2$ &
      $0.2^{b)}$ & $0$ & $0.6$
\\ 2 & $0.71\pm0.27$ &\hspace{3mm} &
      $1.2$ & $0$ & $0$ &
      $0$ & $0$ & $1.2$ 
\\ 3 & $-2.72\pm1.12$ &\hspace{3mm}  &
      $\hspace{-3.3mm}-3.6$ & $0$ & $0.6$ &
      $0$ & $0$ & $-3.0$ 
\\ 4 & $-0.3\pm0.5\;\;$ &\hspace{3mm}  &
      $0$ & $0$ & $\hspace{-3.3mm}-0.5$ &
      $0.5^{b)}$ & $0$ & $0.0$ 
\\ 5 & $0.91\pm0.15$ & \hspace{3mm}  &
      $0$ & $0$ & $1.4^{a)}$ &
      $0$ & $0$ & $1.4$
\\ 6 & $-0.2\pm0.3\;\;$ &\hspace{3mm}   &
      $0$ & $0$ & $\hspace{-3.3mm}-0.3$ &
      $0.3^{b)}$ & $0$ & $0.0$
\\ 7 & $-0.32\pm0.15$ &\hspace{3mm}    &
      $0$ & $0$ & $0$ &
      $0$ & $-0.3$ & $-0.3$
\\ 8 & $0.62\pm0.2\;\;$ & \hspace{3mm}     &
      $0$ & $0$ & $0.9^{a)}$ &
      $0$ & $0$ & $0.9$
\\ 9 & $6.9\pm0.7\;\;$ & \hspace{3mm}      &
      $6.9^{a)}$ & $0$ & $0$ &
      $0$ & $0$ & $6.9$
\\ 10 & $-5.5\pm0.7\;\;$ &\hspace{3mm}      &
      $\hspace{-5.3mm}-10.0$ & $4.0$ & $0$ &
      $0$ & $0$ & $-6.0$
\\ \hline
\end{tabular}
\hbox{$\qquad\qquad\qquad\quad$ $^{a)}$ Input. $\qquad$
$^{b)}$ Estimate based on the limit $N_c \to \infty$. $\qquad$}
\noindent
\caption{Contributions of the resonances $V$, $A$, $S$, $S_1$ and $\eta_1$ 
to the constants $L_i^r$ in units of $10^{-3}$.\protect\cite{VMD} The
phenomenological values in the second column are either from
Ref.\protect\cite{GL1,handbook} (entries 4, 6, 9 and 10) or from
Ref.\protect\cite{ABT} (all the remaining ones). } 
\vspace{0.2cm}
\label{tab:vmd}
\end{center}
\end{table}

These numbers give
\be
l_6^r(M_\rho) \sim -13.3 \times 10^{-3} \co
\label{l6_rho}
\ee
whereas if we extract it from the measured value of the charge radius, we
obtain:
\be
l_6^r(M_\rho) = -(13.5 \pm 2.5)\times 10^{-3} \co
\ee
in very good agreement with (\ref{l6_rho}). This is not a big surprise:
we are simply comparing the experimental determination of the charge radius
with the contribution that the $\rho$ gives to it, and it is well known that
the latter is largely dominant. It is however illuminating on the physical
meaning of the low-energy constants in the chiral expansion. Physically,
these constants must encode the effects of the physics which occurs above
the Goldstone-boson scale, and which does not appear explicitly in the
effective Lagrangian. Intuitively one would expect that the resonances
are the most important ``high-energy'' phenomenon that is neglected
here. A simple estimate of their role should give account of at least the
order of magnitude of these constants.

Actually a careful analysis of the contributions of all the lowest-lying
resonances to the low-energy constants shows that they practically
saturate the experimentally determined values.\cite{VMD}
This is clearly seen from Table \ref{tab:vmd}.

\subsection{Exact solution of the dispersion relation}
If we assume that only the $\pi \pi$ channel is open all the way up to
infinity, then the phase of the form factor on the cut is everywhere the
$\pi \pi$ phase shift $\delta_0^0$. In this approximation, we can pose the
following mathematical problem: suppose that
\begin{enumerate}
\item
$F(t)$ is an analytic function of $t$ in the whole complex plane, with the
exception of a cut for $4 M_\pi^2\leq t < \infty$;
\item
approaching the real axis from above, $e^{-i\delta(t)}F(t)$ is real on the
real axis, where $\delta(t)$ is a known function.
\end{enumerate}
Can $F(t)$ be determined in general?
The solution to this problem is due to Omn\`es,\cite{omnes} and reads as
follows: 
\be
F(t)=P(t)\Omega(t)=P(t)\exp\left\{{t \over \pi} \int_{4 M_\pi^2}^\infty
  {dt' \over t'} {\delta(t') \over t'-t} \right\} \co 
\label{omnes}
\ee
where $P(t)$ is a generic polynomial, and $\Omega(t)$ is called the Omn\`es
function. $P(t)$ can only be constrained by the 
behaviour of the function $F(t)$ at infinity: if we know the behaviour of
the phase $\delta$, and also that of $F(t)$, we can fix at least the degree
of the polynomial $P(t)$. Then, depending on the degree of this polynomial,
we will need a number of inputs to fix its coefficients.
Whatever the degree of the polynomial, the relevant conclusion here is that
if we know the phase, we only need to determine a few constants to fix
the function $F(t)$ completely.

Can we use this information and improve the chiral representation of the
form factor combining it with the Omn\`es representation?

In the chiral representation at order $p^4$ we have seen that there appears
the Omn\`es function expanded to next-to-leading order:
\be
\Gab_{\mbox{\tiny{CHPT}}}^{(2)}(t)=1+b^{(0)}t +\Delta^{(2)}(t)\co \qquad 
\Delta^{(2)}(t)={t\over \pi} \int_{4 M_\pi^2}^{\infty}{dt' \over t'} {
  \delta^{(2)}(t') \over t'-t} \fs
\ee
Indeed the above expression can be seen as a chiral expansion of the Omn\`es
solution (\ref{omnes}):
\bea
\Gab(t)&=&\left(1+bt\right)\Omega(t) \sim \left(1+ [b^{(0)}+ O(M^2)] t \right)
\left[1+\Delta^{(2)}(t) +\cO(p^4) \right] \nn
&=&1+b^{(0)}t +\Delta^{(2)}(t)+\cO(p^4) \co
\eea
and we could try to reconstruct the full Omn\`es solution from its
expansion. An obvious improvement would be, for example:
\be
\Gab^{{\alpha}}(t)=(1+b^{(2)}t)e^{\Delta^{(2)}(t)} \co
\ee
where we have simply exponentiated the dispersive integral over the $\pi
\pi$ phase; or we could use the chiral phase at next-to-leading 
order\cite{GL1} in the exponentiated dispersive integral:
\be
\Gab^{{\beta}}(t)=(1+b^{(2)}t)e^{\Delta^{(4)}(t)}\; ;
\ee
or, if we have a good phenomenological representation for the phase, we
could use this to calculate the dispersive integral:
\be
\Gab^{{\gamma}}(t)=(1+b^{(2)}t)e^{\Delta_{\mbox{\tiny{phys}}}(t)} \fs
\label{gamma}
\ee
Also, in calculating the dispersive integral, we could in principle choose
whether to use a cut-off or to extend the integration up to
infinity. However, given the definition of the Omn\`es function
(\ref{omnes}), the number of subtractions is not sufficient to make the
integral convergent, even for the $p^2$ phase: with a once-subtracted
Omn\`es function using a cut-off is mandatory.

Finally, we have to mention another possible degree of freedom in combining
the chiral and the Omn\`es representation: the choice of the subtraction
point. In the present case there seems to be no choice, because we know
that the form factor at $t=0$ must be equal to 1. It is then natural to
have both the polynomial and the Omn\`es function equal to 1 at
$t=0$. This choice fixes the subtraction point of the Omn\`es function. If
the information on the value of the form factor at zero were missing, we
could as well have chosen a different subtraction point of order $M_\pi^2$,
without changing the chiral counting of the final result.

\begin{figure}[t]
\begin{center}
       \setlength{\unitlength}{1truecm}
       \begin{picture}(10.,10.)
       \epsfxsize 10.  true cm
       \epsffile{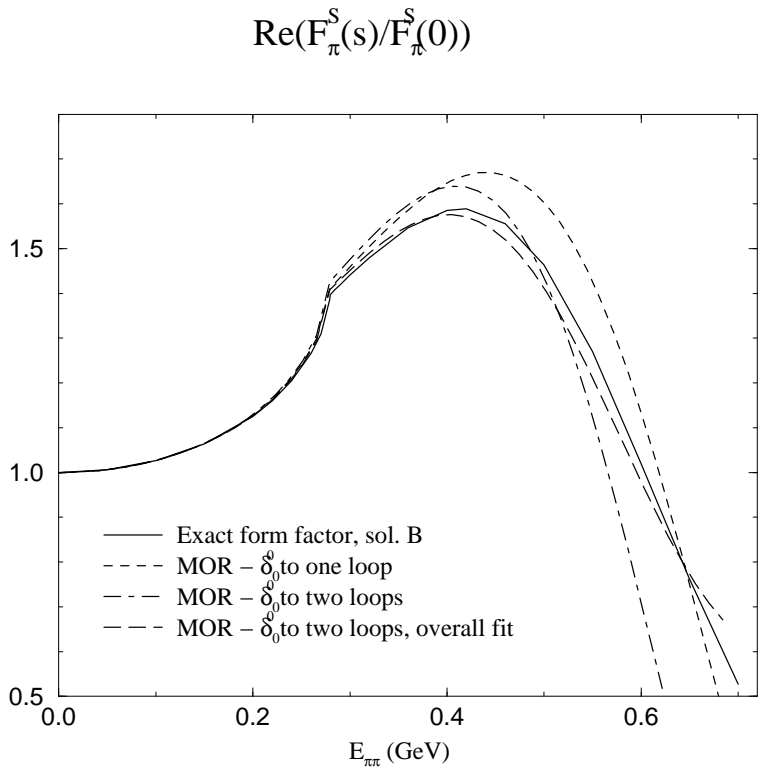}
       \end{picture}
\caption{\label{fig:MOR} Comparison of the physical scalar form factor
  (solid line) with various ``Modified Omn\`es representations'' (MOR). The
  dashed and dot-dashed lines show the difference between the use of the
  phase to one and two loops, whereas the long-dashed line shows that by
  tuning some of the low-energy constants, the MOR can actually reproduce
  the form factor in a wide range of energy.}
\end{center}
\end{figure}

There are various ways to combine the chiral and the Omn\`es
representations. In all cases we improve the chiral representation since we
are able to sum up higher-order terms that are fixed by unitarity. On the
other hand the difference between one way and the other to implement this
improvement is only one order higher than the chiral representation we
started with. We can claim a real improvement only if we can show that this
arbitrariness can be constrained by physical arguments, and that its effect
is numerically small.
For example, If we had a good phenomenological representation for the phase,
with very small uncertainties, it would certainly be preferable to use the
representation $\Gab^{\gamma}(t)$ rather than $\Gab^{\alpha}(t)$, or
$\Gab^{\beta}(t)$. We have already commented on the choice of the
subtraction point: that is another example of an important information that
helps in reducing the arbitrariness of this unitarization procedure.
In the present case the only degree of freedom is in the choice of
the linear term of the polynomial: Eq. (\ref{gamma}) shows that combining
the chiral and the Omn\`es representations amounts to fixing this
subtraction constant with CHPT.

In the recent literature there are various examples of the application of
such procedures: one combines the solution of a dispersion relation (be it
exact, as in the case of the form factor, or numerical, in other cases)
with the chiral representation, and uses the latter only to determine the
subtraction constant. The advantage of such a procedure is clear: the
dispersive integrals, which extend all the way up to infinity, are
certainly dominated by the low-energy region, but also receive sizeable
contributions from the intermediate-energy region, between 0.5 and 1
GeV. Using the chiral representation to evaluate the contribution from that
region is certainly not the best one can do. If one has better information
on imaginary parts, or phases, in that region, one should definitely use
it. As we have discussed here, this procedure is not free from
arbitrariness, and all care should be used to constrain this to a
minimum. The interested reader is referred to the original
articles\cite{disp&chiral} for a more detailed treatment of the case of the
pion form factors. A review of the marriage between CHPT and dispersion
relations can be found in Ref.\cite{donoghue}

\subsection{Concluding remarks on lectures 2 and 3}
In these two lectures we have analysed in detail the various contributions
entering the next-to-leading order calculation of a typical amplitude in
CHPT. To make the analysis concrete we have considered the scalar form
factor. 

We have seen that:
\begin{enumerate}
\item The renormalization procedure is straightforward -- indeed not very
  different from the case of a renormalizable field theory.
\item The loop integrals generate infrared divergences that are physical:
  the ef\-fec\-tive-Lagrangian method is a systematic way to calculate them.
\item The finite, analytically non-trivial part of the loop integrals is
  dictated by the properties of unitarity and analyticity. As usual in
  quantum field theory these properties are built in, and automatically
  respected in loop calculations.
\item The finite part of the counterterms can be expressed in terms of sum
  rules and in most cases these are saturated by the
  lowest-lying resonances. The low-energy constants have a very clear
  physical meaning, as they embody the effect of the high-energy degrees
  of freedom that are not explicitly
  considered in the effective Lagrangian framework 
\end{enumerate}

\section{Non-leptonic weak interactions}

\subsection{Partonic $|\Delta S| =1$ effective  Hamiltonian}

As discussed in the first lecture, the chiral realization of $Z(v,a,s,p)$
allows us to calculate not only transition amplitudes of pure QCD, but also
Green functions of weak and electromagnetic processes with external gauge
fields, such as semileptonic kaon and pion decays. $Z(v,a,s,p)$ is not
sufficient, however, to describe non-leptonic weak transitions, such as $K\to
2 \pi$ and $K\to 3 \pi$ decays.  In this kind of processes the $W$ boson is
coupled to two quark currents and cannot be treated as an external
field. The strong-interaction effects that renormalize the non-leptonic
weak transition cannot be trivially factorized and should be evaluated up
to distances of $\cO(1/M_W)$.

The simplest strategy to describe non-leptonic weak interactions is based
on a twofold EQFT approach. The first step, performed within perturbative
QCD, is the construction of the partonic $|\Delta S| =1$ effective
Hamiltonian, or the expansion of the non-local product of weak currents
into a series of local partonic operators renormalized at a scale $\mu\gsim
1$~GeV (see e.g.~Buchalla {\em et al.}\cite{BBL} for a comprehensive
review).  This first step, based on Wilson's Operator Product
Expansion,\cite{WilsonOPE} allows us to encode in appropriate coefficients
the sizeable QCD corrections that renormalize the weak interaction at
short distances, from $M_W$ down to the lowest scale where perturbative QCD
can still be applied,  considerably simplifying  the original problem.

\begin{figure}[t]
    \begin{center}
       \setlength{\unitlength}{1truecm}
       \begin{picture}(6.0,3.5)
       \epsfxsize 6.  true cm
       \epsfysize 3.5 true cm
       \epsffile{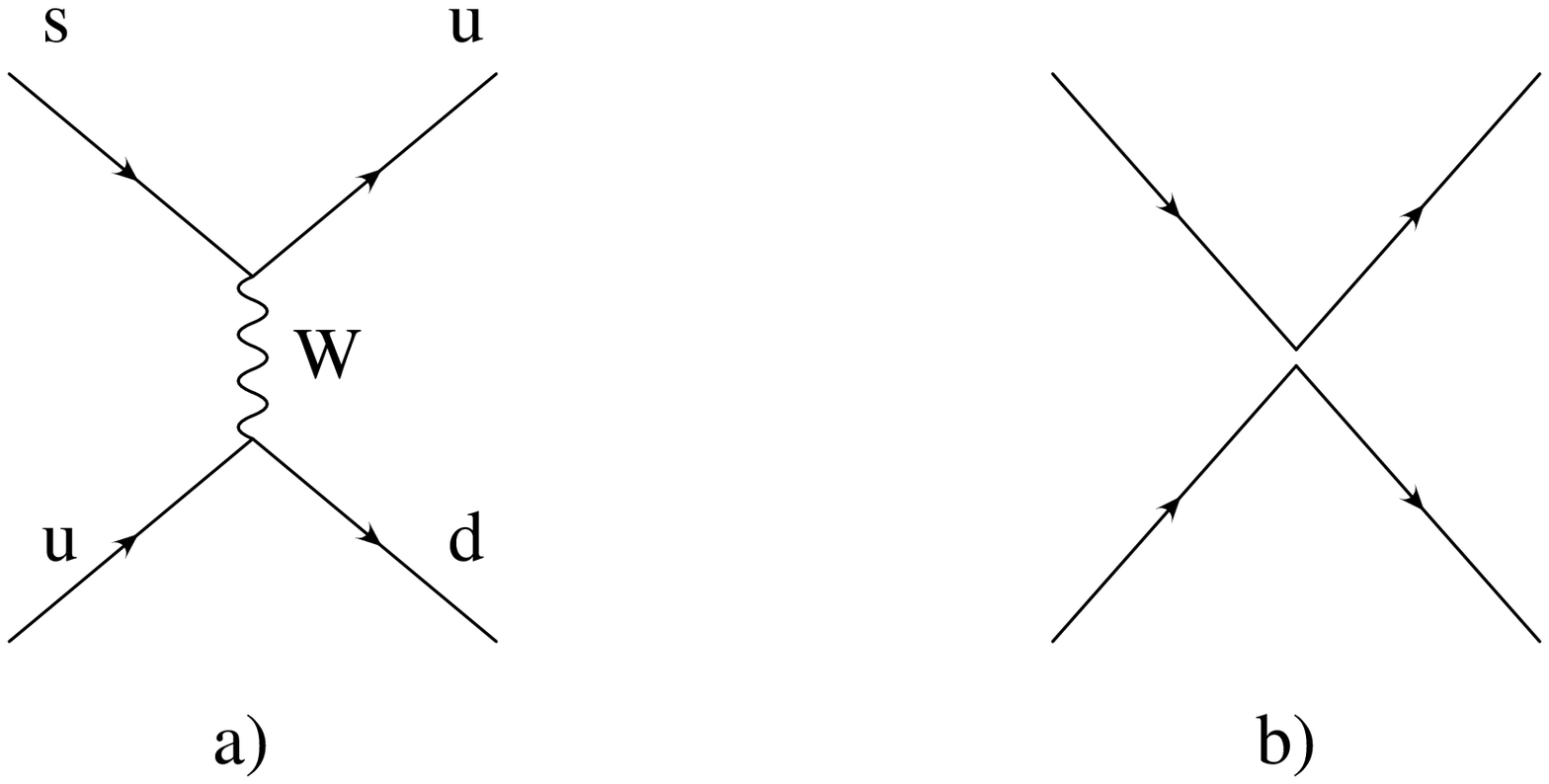}
       \end{picture}
    \end{center}
    \caption{a) Tree-level Feynman diagram 
for $|\Delta S|=1$ transitions, at the lowest order in
$G_F$ and without strong-interaction corrections; b) 
the same diagram in the effective theory.}
    \protect\label{fig:4_1}
    \begin{center}
       \setlength{\unitlength}{1truecm}
       \begin{picture}(10.0,2.5)
       \epsfxsize 10.  true  cm
       \epsfysize 2.5  true cm
       \epsffile{lnfss_f2.eps}
       \end{picture}
       \centerline{a)}
       \begin{picture}(10.0,2.5)
       \epsfxsize 10.  true  cm
       \epsfysize 2.5  true cm
       \epsffile{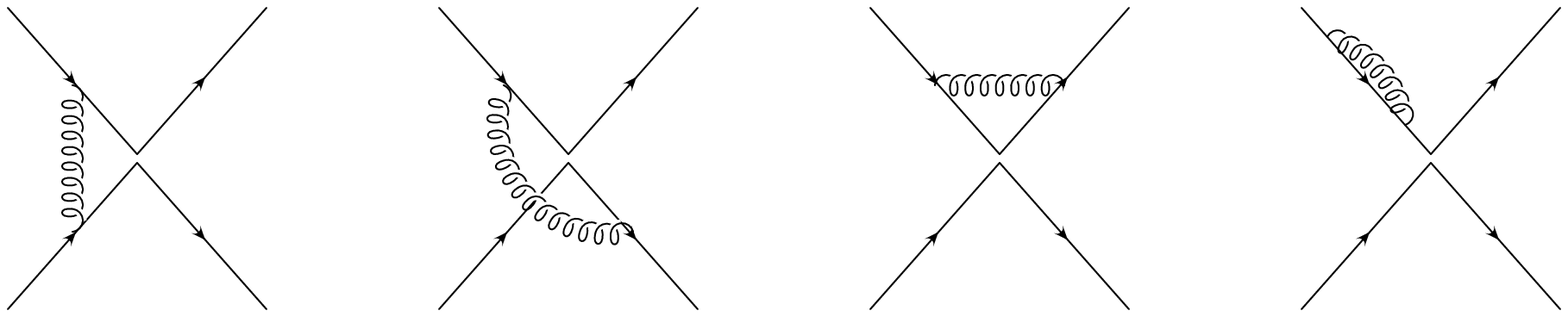}
       \end{picture}
       \centerline{b)}
    \end{center}
    \caption{Leading-order QCD corrections to the diagram in 
      Fig.~\protect\ref{fig:4_1}: 
      a) in the full theory; b) in the effective theory.}
    \protect\label{fig:4_2}
\end{figure}

Denoting by  $J_\mu(x)$ the charged weak current and by $D^{\mu\nu}_W(x,M_W)$ 
the $W$ propagator in  spatial coordinates, the partonic $|\Delta S| =1$ 
effective Hamiltonian at $\cO(G_F)$, 
\beq
{\cal H}_{eff}^{|\Delta S|=1} = {G_F \over  \sqrt{2}}   V_{us}^* V_{ud} 
 \sum_i C_i(\mu) Q_i(\mu) + {\rm h.c.}~,
\eeq
can formally be defined by the following equation
\beqa
\cT( I \to F ) &=& { g^2 \over 8 }\int {\rm d}^4 x
D_W^{\mu\nu}(x,M_W) \bra{F}|T\left( J_\mu(x)J\da_\nu(0) \right) \ket{I} + \cO(G_F) \no\\
&=& - {G_F \over  \sqrt{2}} V_{us}^* V_{ud} 
 \sum_i C_i(\mu) \bra{F}| Q_i(\mu) \ket{I} + \cO(G_F)~.
\label{ope}
\eeqa
As can easily be verified by expanding the $W$ propagator in 
Fig.~\ref{fig:4_1}, 
in the absence of QCD interactions the sum in the right-hand side of Eq.~(\ref{ope})
would be restricted to only two operators, 
\beq
Q_\pm= 2 \left[~\bar{s}_L \gamma^\mu u_L\bar{u}_L \gamma^\mu d_L~
\pm ~\bar{s}_L \gamma^\mu d_L\bar{u}_L \gamma^\mu u_L~
\right]~,
\label{Oqpm}
\eeq
whose (Wilson) coefficients are given by 
\beq
C_+ = C_- = 1~.
\label{eq:Cpm0}
\eeq

\noindent
The situation becomes more complicated when the QCD corrections in
Fig.~\ref{fig:4_2} are taken into account. In order to evaluate these
effects within perturbative QCD it is assumed that external quarks carry
large momenta. Moreover, since  some diagrams
develop artificial divergences (absent in the real case)
within the effective theory, it is necessary
to regularize the effective operators by means of the Wilson coefficients,
introducing an arbitrary renormalization scale $\mu$.  The requirement that
the product $C_i(\mu) \times Q_i(\mu)$, and thus all physical observables,
be independent of $\mu$, fixes  the renormalization-group
evolution of the coefficients $C_i$ as a function of $\mu$ unambiguously.  
The initial
value of the $C_i$ is fixed by matching the effective theory to the full
one at high scales [i.e.~for external momenta of $\cO(M_W)$], where the
perturbative calculation is rapidly convergent. Then, using
renormalization-group equations the $C_i$ are evolved down to a scale of
$\cO(1~{\rm GeV})$, as close as possible to the physical scale of the
process (the kaon mass), but still high enough to trust perturbative QCD.

As a result of this procedure, all leading logarithms,  
i.e.~all the terms of $\cO[\alpha^n_s \log(M_W/\mu)^n]$,
are encoded in the Wilson coefficients. 
At this level of accuracy the set of effective operators 
is still the one in Eq.~(\ref{Oqpm}); however, their 
coefficients have changed to\cite{AM}
\beqa
C_+(\mu)  &=& C_+(M_W) \left(\frac{\alpha_s(M_W)}{\alpha_s(\mu)}\right)^{\frac{1}{2\beta_0}} =  \left(\frac{\alpha_s(M_W)}{\alpha_s(\mu)}\right)^{\frac{1}{2\beta_0}}~, 
 \no \\
C_-(\mu)  &=& C_-(M_W) \left(\frac{\alpha_s(M_W)}{\alpha_s(\mu)}\right)^{-\frac{1}{\beta_0}} =  \left(\frac{\alpha_s(M_W)}{\alpha_s(\mu)}\right)^{-\frac{1}{\beta_0}} ~,
\label{eq:Cpm1}
\eeqa
where $\beta_0 = (33 -2 N_F/12)$ and  $N_F$ denote the number of dynamical quarks. 
The weight of the operator $Q_-$ is therefore enhanced, whereas that of 
$Q_+$ is decreased. 

This program can be iterated at the next-to-leading order,
resumming also subleading logarithms. At this level
the number of terms increases, with the inclusion of the 
so-called penguin operators. Finally, when also the 
small electroweak $\cO(e^2G_F)$ operators relevant to 
$CP$-violation studies are taken into account, the full basis 
of ${\cal H}_{eff}^{|\Delta S|=1}$ below the charm 
threshold contains, in addition to $Q_\pm$, the following set 
of operators:\cite{BBL}
\beqa
Q_{3(5)} &=& 4~\bar{s}_L^\alpha \gamma^\mu d_L^\alpha~\sum_{q=u,d,s}
         ~\bar{q}_{L(R)}^\beta  \gamma^\mu q_{L(R)}^\beta~,  \no\\
Q_{4(6)} &=& 4~\bar{s}_L^\alpha \gamma^\mu d_L^\beta~\sum_{q=u,d,s}
         ~\bar{q}_{L(R)}^\beta  \gamma^\mu q_{L(R)}^\alpha~,  \no\\
Q_{7(9)} &=& 6~\bar{s}_L^\alpha \gamma^\mu d_L^\alpha~\sum_{q=u,d,s}
         e_q~\bar{q}_{R(L)}^\beta  \gamma^\mu q_{R(L)}^\beta~,  \no\\
Q_{8(10)}&=& 6~\bar{s}_L^\alpha \gamma^\mu d_L^\beta~\sum_{q=u,d,s}
         e_q~\bar{q}_{R(L)}^\beta  \gamma^\mu q_{R(L)}^\alpha~, 
\label{ciubase}
\eeqa
where $\alpha$ and $\beta$ are colour indices and $e_q$ 
denotes the electric charge of the quark $q$.

Thanks to the resummation of the subleading logarithms, the coefficients of
the $Q_i$ are known with high accuracy.\cite{BBL,Ciuchini} According to
Eq.~(\ref{ope}), the remaining problem to be addressed in order to
calculate the transition amplitudes of physical processes is the evaluation
of the hadronic matrix elements of the $Q_i$.  This issue is the goal of
the second EQFT construction.

\subsection{Lowest-order non-leptonic $|\Delta S|=1$ chiral Lagrangian}

Following the basic principles of CHPT, in order to compute non-leptonic
weak transitions of pseudoscalar mesons we need to construct the chiral
realization of ${\cal H}_{eff}^{|\Delta S|=1}$, or we need to consider the
most general Lagrangian, written in terms of pseudo-Goldstone fields,
transforming as ${\cal H}_{eff}^{|\Delta S|=1}$ under $SU(3)_L\times
SU(3)_R$.
                    
The operators of 
Eq.~(\ref{ciubase}) transform linearly under 
$SU(3)_L\times SU(3)_R$ in the following way:
\beq \ba{lcl}
Q_-, Q_3, Q_4, Q_5, Q_6    & \qquad & (8_L,1_R)~,  \\    
Q_+, Q_9, Q_{10}           &        & (8_L,1_R) + (27_L,1_R)~,   \\   
Q_7, Q_8                   &        & (8_L,8_R)~.  \ea
\label{trasfOi} 
\eeq
Analogously to the case of light-quark masses,
chiral operators transforming like the $Q_i$ 
can be built by introducing appropriate external sources. 
As an example, in order to build the  $(8_L,1_R)$ operators
we introduce the  source  
\beq
{\hat \lambda}  \Gto   g_{_L} {\hat \lambda}  g_{_L}^{-1}
\eeq
and we consider all possible operators invariant 
under $G$ and linear  in $\hat \lambda$.
Then, fixing the source to the constant value 
\beq
{\hat \lambda}  \to \lambda={1\over 2}(\lambda_6-i\lambda_7) = 
\left( \ba{ccc} 0 & 0 & 0 \\ 0 & 0 & 0 \\ 0 & 1 & 0 \ea \right)~,
\eeq
we automatically select the $|\Delta S|=1$ component 
of all possible $(8_L,1_R)$ terms. For 
$(27_L,1_R)$  and $(8_L,8_R)$ operators the procedure is 
very similar, the only change is the source structure. 

Interestingly, within each group of transformations
there is only one chiral realization at the lowest order. 
These are given by:\cite{Cronin2,GRW}
\beq 
\ba{lcll}
W^{(2)}_8    = \la \lambda L_\mu L^\mu \ra  & \qquad & (8_L,1_R) \qquad 
& \cO(p^2)~, \\
W^{(2)}_{27} = (L_\mu)_{23} (L^\mu)_{11} 
   + {2\over 3}(L_\mu)_{21} (L^\mu)_{13}    &  & (27_L,1_R)
& \cO(p^2)~,  \\
W^{(0)}_{\underline{8}} = F^2 \la \lambda U\da Q U \ra  &   & (8_L,8_R)
& \cO(p^0)~,  \ea
\eeq  
where $L_\mu=u\da u_\mu u$. While the singlets under 
$SU(3)_R$ are at least of $\cO(p^2)$, the lowest-order 
realization of $(8_L,8_R)$ operators starts at $\cO(p^0)$.
This difference does not create a power-counting 
mismatch since at the quark level 
the $(8_L,8_R)$ terms are suppressed by a factor $e^2$ 
with respect to the $SU(3)_R$ singlets.
Thus we can consistently consider the 
chiral realization of the $|\Delta S|=1$ Lagrangian
at $\cO(G_Fp^2e^0)+\cO(G_Fp^0e^2)$:  
\beq
{\cal L}_W^{(2)} =  F^4 \left[  G_8 W_8^{(2)} + G_{27} W_{27}^{(2)} 
+  G_{\underline{8}} W_{\underline{8}}^{(0)}
\right ] + \mbox{\rm h.c.}
\label{lagr2w}
\eeq

The constants $G_i$ appearing in ${\cal L}_W^{(2)}$ are not
fixed by symmetry arguments, apart from the constraint 
$\Im(G_i)=0$, which holds in the limit where $CP$ is an 
exact symmetry. By construction we can only write
\beq
\cT( I \to F ) = 
- {G_F \over  \sqrt{2}} 
V_{us}^* V_{ud}
  \sum_i C_i(\mu) \bra{F}| Q_i(\mu) \ket{I} 
=  F^4 \sum_i G_i  \bra{F}|   W_i^{(2)} \ket{I} +\cO(p^4) \quad
\label{eq:CiGi}
\eeq 
and unfortunately in this equation we have unknown terms on both sides
of the last identity: the $\bra{F}| Q_i(\mu) \ket{I}$ on the left and the
$G_i$ on the right. As we will show below, since the $G_i$ are few, we can
determine them all by comparison with experimental data in $K\to 2 \pi$
decays, i.e.~by measuring $\cT( K \to 2\pi )$. In this case
Eq.~(\ref{eq:CiGi}) lets us fix the matrix element of some combination of
$Q_i(\mu)$ between a kaon and two pions and then, more interestingly, lets us
{\em predict} the matrix element of the same combination of $Q_i(\mu)$ in
other channels, e.g.~between a kaon and three pions. In the future one could
hope to replace, fully or in part, the use of experimental inputs with a
theoretical determination by means of lattice-QCD of some $\bra{F}|
Q_i(\mu) \ket{I}$. Note, however, that in both approaches, either using
lattice QCD or experimental data, the role of CHPT is to relate the matrix
elements of the $Q_i(\mu)$ in different channels and not to predict from
scratch the $\bra{F}| Q_i(\mu) \ket{I}$ in a given channel.

Rough theoretical estimates of the $G_i$ can be obtained 
by using  simplifying assumptions that let us compute 
the $\bra{F}| Q_i(\mu) \ket{I}$ explicitly, although 
with large (and typically uncontrolled) uncertainties. 
These are very useful to understand the order of 
magnitude of the $G_i$. The most natural 
simplifying assumption to estimate the  $\bra{F}| Q_i(\mu) \ket{I}$
is the factorization hypothesis, that can be formally 
justified within QCD in the limit of an infinite number of colours 
($N_C \to \infty$).\cite{BBG} 
According to this hypothesis, the hadronization of 
operators such as $Q_\pm$, with a colour-singlet
(current)$\times$(current) structure, is given
by the product of the corresponding hadronized
currents. Since the hadronization of a colour-singlet 
quark current is completely determined by the coupling 
of the external sources in the strong Lagrangian 
[see Eqs.~(\ref{lqcdvasp})--(\ref{chiralF})],
\beq
\bar{q}^i_L \gamma_\mu q^j_L \to \frac{iF^2}{2}
\left(\partial_\mu U\da  U \right)_{ji}~,
\eeq
this hypothesis leads to a full determination of the 
hadronization of $Q_\pm$:
\beq
 Q_-  \to {F^4 \over 2}  W^{(2)}_8~, \qquad 
 Q_+  \to   F^4  \left[ {3 \over 5}  W^{(2)}_{27} +  
           {1 \over 10} W^{(2)}_{27} \right]~. 
\eeq
In this case there is no trace of the
renormalization scale $\mu$, and indeed it can be 
shown that the anomalous dimensions of $Q_\pm$ 
vanish in the limit $N_C \to \infty$. 
Using Eq.~(\ref{eq:CiGi}) one then finds 
\beqa
 \left. G_8\right|_{\rm fact.}  ~=& - \dis\frac{G_F}{\sqrt{2}} 
V_{us}^* V_{ud} \left( {1\over 2} 
  C_- + {1\over 10} C_+ \right)  ~\stackrel{ _{N_C \to \infty} }{\longrightarrow} 
  &  - \frac{3G_F}{5\sqrt{2}}  V_{us}^* V_{ud}~, \no\\
 \left. G_{27}\right|_{\rm fact.} ~=& - \dis\frac{G_F}{\sqrt{2}} 
V_{us}^* V_{ud}
\left( {3\over 5} C_+ \right) \qquad\qquad~\stackrel{ _{N_C \to \infty} }{\longrightarrow}  &  - \frac{3G_F}{5\sqrt{2}}  V_{us}^* V_{ud}~,
\label{eq:Cpmf}
\eeqa
where in the terms after the arrows we have employed the 
values of $C_\pm$ in Eq.~(\ref{eq:Cpm0}), which are consistent 
with the absence of anomalous dimensions for $Q_\pm$. 
As  can be noted, the normalization of 
${\cal L}_W^{(2)}$ is such that the $G_i$ are 
expected to be $\cO(G_F V_{us} C_j)$, where $C_j$ are 
Wilson coefficients of partonic operators 
transforming as the corresponding chiral realizations.

As anticipated, the $G_i$ can be experimentally determined 
by means of $K\to 2 \pi$ data. As long as we are interested 
in the real parts or the absolute values of the decay amplitudes 
we can neglect $G_{\underline{8}}$. Indeed the Wilson coefficients 
of $(8_L,8_R)$ operators are absolutely negligible with respect 
to the others --  for the imaginary parts this is not the case. 
Introducing the isospin amplitudes $A_{0}$ and $A_{2}$
\beqa 
\cA(\Ko \to \pi^+ \pi^-)&=&A_0e^{i\delta_0}+{1 \over \sqrt{2}} A_2e^{i\delta_2}~, 
\no \\
\cA(\Ko \to \pi^0 \pi^0)&=&A_0e^{i\delta_0}- \sqrt{2} A_2e^{i\delta_2}~,
\label{k2pdec} \label{definA_0A_2}  \\
\cA( K^+ \to \pi^+ \pi^0)&=&{3 \over   2} A_2e^{i\delta_2}~, \no
\eeqa
it is easy to verify that 
\beqa
A_{0}  &=& \sqrt{2} F  \left(M_K^2 -M_\pi^2\right)\left(G_8+{1\over 9} 
G_{27} \right)~, \label{A_01}\\
A_{2} &=&  {10 \over 9} F G_{27} \left(M_K^2-M_\pi^2 \right)~.
\label{A_21}
\eeqa   
The comparison with the experimental data then leads to: 
\beqa
|G_8| &=& 9.1 \times 10^{-6}\ \mbox{\rm GeV}^{-2}~, \label{G_8xx} \\
G_{27}/G_8  &=& 5.7 \times 10^{-2}~.
\eeqa 
Interestingly, the experimental values of $G_{8}$ and $G_{27}$ 
are substantially different from their naive estimates 
in Eq.~(\ref{eq:Cpmf}): the absolute value of  
$G_{8}$ is about 8 times larger, whereas $|G_{27}|$ 
is reduced to 50\%. This phenomenon, known as the
``$\Delta I =1/2$ rule'', is in part explained by
the running of the Wilson coefficients. 
Indeed if in Eq.~(\ref{eq:Cpmf})
one uses the leading-log values of $C_\pm(\mu)$,
with a renormalization scale $\mu\sim 1$~GeV, then the factorized 
estimate of $|G_{8}|$ increases by a factor of about 2 and the 
one of $|G_{27}|$ reduces to 70\%. 
Although this is encouraging, it is clear that there still is a 
large non-perturbative effect hidden in the matrix elements
of four-quark operators, especially the $(8_L,1_R)$ ones.
At the moment the best we can do is to measure this effect 
using $K\to 2\pi$ data.

Once the $G_i$ have been fixed from $K\to 2\pi$, the theory is absolutely
predictive in all the other non-leptonic channels. In Table~\ref{tab:K3p1}
we show the comparison between theory and experiments in $K\to 3\pi$
amplitudes. The latter are classified according to the variation of isospin
($\Delta I=1/2$ or $\Delta I=3/2$) and to the dependence from Dalitz-plot
variables, as summarized in Table~\ref{tab:K3p2}.  In agreement with naive
power counting, the discrepancy between lowest-order chiral predictions and
data, within the dominant amplitudes, turns out to be around $30\%$.  On
the other hand, $\cO(p^2)$ operators have not enough derivatives to produce
a non-vanishing result for the suppressed quadratic slopes.  The comparison
between data and lowest-order CHPT predictions is much more satisfactory in
the case of $K\to 2\pi\gamma$ decays, where the $\cO(p^2)$ amplitudes
coincide with the bremsstrahlung from $K\to 2\pi$, which is known to be
largely dominant.\cite{DAI}

\begin{table}[t]
\begin{center}
\begin{tabular}{|c||c|c|c|}\hline
Amplitude & $\qquad \cO(p^2) \qquad$ & $\qquad \cO(p^4) \qquad$
 & Experimental fit \ \     \\ \hline\hline
$\alpha_1$& $74.0$   & input    & $ 91.71\pm0.32$  \\ \hline
$\beta_1$ & $-16.5$  & input    & $-25.68\pm0.27$ \\ \hline
$\zeta_1$ & $-$      & $-0.47\pm 0.18$   & $-0.47\pm 0.15$ \\ \hline
$\xi_1$   & $-$      & $-1.58\pm 0.19$   & $-1.51\pm 0.30$ \\ \hline
$\alpha_3$& $-4.1$   & input    & $-7.36\pm 0.47$  \\ \hline
$\beta_3$ & $-1.0$   & input    & $-2.43\pm 0.41$ \\ \hline
$\gamma_3$& $1.8$    & input    & $2.26 \pm 0.23$  \\ \hline
$\zeta_3$ & $-$      & $-0.011\pm 0.006$  & $-0.21\pm 0.08$ \\ \hline
$\xi_3$  & $-$      &  $0.092\pm 0.030$  & $-0.12\pm 0.17$ \\ \hline
$\xi'_3$   & $-$      & $-0.033\pm 0.077$  & $-0.21\pm 0.51$ \\ \hline
\end{tabular}
\caption{Experimental results vs.
CHPT predictions in $K\to3\pi$ decays.\protect\cite{KMWlett}
\protect\label{tab:K3p1}}
\end{center}
\end{table}

\begin{table}
\begin{center}
\begin{tabular}{|l|l|l|c|}\hline
                & $\Delta I=1/2$      & $\Delta I=3/2$ & Phases        \\ \hline \hline 
Constant terms  & \ \  1 ($\alpha_1$)   &   \ \ 1   ($\alpha_3$)            &   1 \\  \hline
Linear slopes   & \ \  1  ($\beta_1$ )  &   \ \ 2   ($\beta_3,\gamma_3$ )   &   3  \\ \hline
Quadratic slopes      &  \ \ 2 ($\zeta_1,\xi_1$)  &  \ \  3 ($\zeta_3,\xi_3,\xi'_3$)
                &   -      \\ \hline\hline
$\cO(p^4)$  free parameters  &  \ \ 2   & \ \ 3 & - \\ \hline
\end{tabular}
\caption{Number of independent isospin amp\-li\-tu\-des vs.
$\cO(p^4)$ free parameters in $K\to3\pi$ decays.\protect\label{tab:K3p2}}
\end{center}
\end{table}
    
\subsection{$\cO(p^4)$ counterterms}
\label{subsez:LW4}

In order to obtain a more refined description of experimental data, 
able to include absorptive effects and subleading amplitudes
(such as $K \to 3 \pi$ quadratic slopes and  $K \to \pi\pi\gamma$ 
direct-emission terms), 
it is necessary to go beyond the lowest order in the chiral 
expansion. Since we are interested only in contributions 
of order $G_F$, we can proceed analogously to the case 
of strong interactions with the simple substitution
\beqa
{\cal L}_S^{(2)} &\to &{\cal L}_S^{(2)}+ {\cal L}_W^{(2)}~,\no\\
{\cal L}_S^{(4)} &\to &{\cal L}_S^{(4)}+ {\cal L}_W^{(4)}~.
\eeqa 
Here ${\cal L}_W^{(4)}$ denotes the most general $ \cO(p^4)$ 
Lagrangian transforming like ${\cal L}_W^{(2)}$ under chiral rotations 
and thus able to absorb all the one-loop divergences generated 
by ${\cal L}_S^{(2)}\times{\cal L}_W^{(2)}$.

The $\cO(p^4)$ operators transforming like $(8_L,1_R)$ and $(27_L,1_R)$
were classified for the first time by Kambor {\em et
al.},\cite{KMWnucl} about 10 years ago, whereas the $\cO(e^2p^2)$ terms
transforming as $(8_L,8_R)$ have been analysed only very
recently.\cite{Cirigliano} The overall picture is certainly worse than in
the case of strong interactions, since the number of independent operators
is much larger: already within the dominant $(8_L,1_R)$ sector there are 37
independent terms.\cite{EckerWyler} Nonetheless the theory still has a
considerable predictive power, since out of these terms only few
combinations appear in observable processes.\cite{DAI,EckerWyler} For
instance, within $K \to 3 \pi$ decays the number of independent counterterm
combinations coincides with the number of leading (constant and linear)
amplitudes. Thus a fit of the $\cO(p^4)$ parameters from the leading
amplitudes leads to unambiguous predictions for the quadratic slopes, as
shown in Tables~\ref{tab:K3p1} and \ref{tab:K3p2}.

\begin{figure}[t]
    \begin{center}
       \setlength{\unitlength}{1truecm}
       \begin{picture}(10.0,7.0)
       \epsfxsize 10. true  cm
       \epsfysize 7.  true cm
       \epsffile{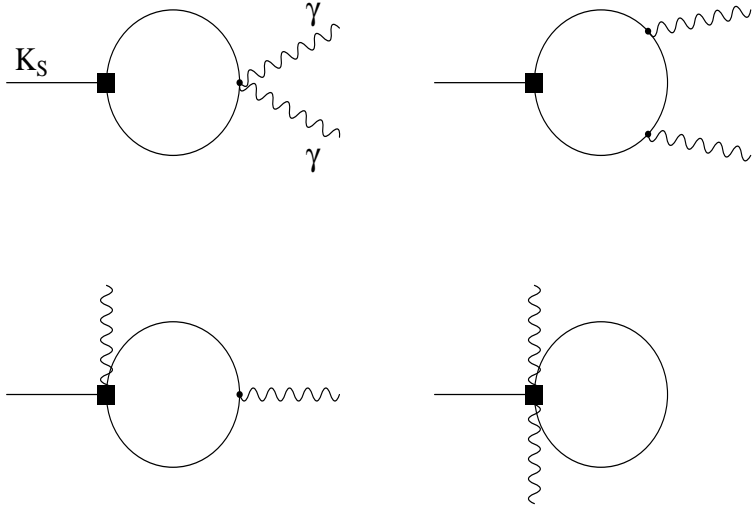}
       \end{picture}
    \end{center}
    \caption{$\cO(p^4)$ diagrams for the transition $K_S \to \gamma
              \gamma$. The black box denotes the weak vertex.}
    \label{fig:kggloop}
\end{figure}

There are even cases of physical processes that receive 
$\cO(p^4)$ contributions only from loop diagrams and 
not from counterterms, like the $K_S \to \gamma \gamma$ decay.
The leading contributions to  $\cA(K_S \to \gamma \gamma)$
are the  loop diagrams in Fig.~\ref{fig:kggloop}. 
Since there are no local contributions to this transition
at $\cO(p^4)$, the loop amplitude turns out to be finite\cite{DE} 
and leads to the following parameter-free result:
\beq
\cB(K_S\to \gamma\gamma)^{(4)} = 2.1 \times 10^{-6}~.
\eeq
The above prediction turns out to be
in excellent agreement with the experimental data, 
$\cB(K_S\to \gamma\gamma)^{\rm exp.} = (2.5 \pm 0.5)\times 10^{-6}$,
providing a very significant test of the quantum nature 
of this effective field theory.

\subsection{Beyond $\cO(p^4)$}

Another process that receives
$\cO(p^4)$ contributions only from loop diagrams is 
$K_L \to \pi^0 \gamma\gamma$.
The diagrams describing this transition are 
very similar to those in Fig.~\ref{fig:kggloop};
the only difference is an external pion leg 
attached to the weak vertex. 
In this case, however, the $\cO(p^4)$ parameter-free
prediction of the branching ratio 
\beq
\cB(K_L\to \pi^0 \gamma\gamma)^{(4)} = 0.6 \times 10^{-6}
\label{Klwidht}
\eeq
turns out to be in bad agreement with the experimental finding: 
$\cB(K_L \to \gamma\gamma)^{\rm exp.} = (1.7 \pm 0.1)\times 10^{-6}$.
What is the reason for the big difference between 
$K_S\to \gamma\gamma$ and $K_L\to \pi^0 \gamma\gamma$?
The answer to this question can be traced back to what 
we learned in lecture 3. 

The first difference is related to the absorptive 
parts of the amplitudes. Since the coupling $G_8$ 
has been fitted from $K_S \to 2 \pi$, 
the imaginary part in $K_S\to \gamma\gamma$, computed at the one-loop level,
perfectly reproduces the absorptive contribution due to the  
$K_S \to 2 \pi$ intermediate state. On the contrary,  
because of the underestimate of $K\to 3\pi$ amplitudes at 
$\cO(p^2)$ (see Table~\ref{tab:K3p1}), the one-loop imaginary part of 
$K_L\to \pi^0 \gamma\gamma$ underestimates the real 
absorptive contribution to this channel by 20\%--30\%.
The second important difference between $K_S\to \gamma\gamma$ and 
$K_L\to \pi^0 \gamma\gamma$ is induced by resonance contributions.
Vector and axial-vector  mesons, which are known to produce sizeable 
effects in the strong sector (see Table~\ref{tab:vmd}), 
do not appear in $K_S\to \gamma\gamma$ but can affect the
$K_L\to \pi^0 \gamma\gamma$ amplitude. In the latter case one 
can therefore expect a sizeable local $\cO(p^6)$ counterterm,
encoding the contribution induced by vector-meson exchange. 

These two effects shows that, contrary to $K_S\to \gamma\gamma$, a good
description of $K_L\to \pi^0 \gamma\gamma$ requires the inclusion of
$\cO(p^6)$ terms.\cite{DAI} A similar situation emerges in many other
channels and can be interpreted as a general rule: unitarity corrections
induced by pion loops and vector meson resonances provide a useful guide
toward the estimate of the most significant higher-order corrections.

\section{Conclusions}
Solving strong interactions at low energy is a very difficult task, which
so far nobody has been able to complete with analytical methods. On the
other hand, at low energy, the approximate chiral symmetry does impose
severe constraints on Green functions in the form of Ward identities. In
studying the phenomenology of strong interactions at low energies it is
extremely useful to take into account these symmetry constraints.
The effective Lagrangian method is a tool to derive these constraints in an
automatic manner -- at the same time respecting also the general properties
of unitarity and analyticity. 

In this series of lectures we have introduced the basic concepts and
technical tools for using this method. The last lecture was devoted to
the study of a few examples of physically interesting decay channels of the
kaons. In recent years a rich experimental activity at various facilities
around the world has started, with the aim to study, with very high
accuracy, all possible decay modes of the kaons, in some cases measuring
extremely small branching ratios. 
The use of chiral perturbation theory for analysing this rich phenomenology
is essential: it is not by chance that this lecture series was given at 
the Frascati Laboratories, which host one of the main world facilities for this
kind of physics. 

Kaon physics is certainly not the only subfield of strong interactions
where CHPT is useful and is applied successfully. A subject that we have
not touched at all in these lectures, but where the application of CHPT is
nowadays routine is the physics of baryons, going from the ``easy''
one-baryon sector ($\pi N$ scattering, $\pi N \to \pi \pi N$, photo- and
electro-production reactions, etc.) to the more complicated sectors
involving two or more baryons.

In perspective, we believe that CHPT will become more and more used
in connection with lattice calculations: as we have seen in the few
examples of calculations discussed here, the dependence on the
quark masses is always explicit. Whenever a quantity is calculated on the
lattice one can use CHPT to determine the explicit dependence on the quark
masses and make the extrapolation to the physical quark masses in a
controlled manner. The ``simple procedure'' we described here hides in
fact some serious technical difficulties -- conceptually, however, it is
obvious that merging the two methods offers clear advantages, and we will see
more and more of them in the future. The technical difficulties are only
welcome, because it means there will be work for theorists!

Let us finally conclude by mentioning that this method is also applicable
to very different physical systems. In fact it can be applied to any
physical system in which spontaneous symmetry breaking occurs, and which
has an energy gap such that, at low energy, only the Goldstone modes
dominate the physics. We find very interesting, for example, the
applications to magnetic systems -- in some cases, precisely the same
effective Lagrangian that we have constructed for pions can be used to
describe the behaviour of one of these.\cite{hofmann}

\section*{Acknowledgements}
%
%
We thank Lia Pancheri for the invitation to present these 
lectures, and for her perseverance and patience in waiting for 
their written version. This work was supported in
part by the Schweizerische Nationalfonds and by 
the EC-TMR Program N.~ERBFMRX-CT980169 (EURODA$\Phi$NE).


\end{document}